\documentclass[letterpaper, twocolumn]{article}

\usepackage[T1]{fontenc}

\usepackage{geometry}
\usepackage{setspace}
\usepackage{lipsum}

\usepackage[style=chem-acs, backend=biber, doi=true, articletitle=true]{biblatex}
\DeclareSourcemap{
  \maps[datatype=bibtex]{
    \map{
      \step[fieldsource=shortjournal, final]
      \step[fieldset=journaltitle, origfieldval]
    }
  }
}

\usepackage{graphicx}
\usepackage{subcaption}
\usepackage{float}
\newfloat{scheme}{htbp}{los}
\floatname{scheme}{Scheme}
\floatname{chart}{Chart}
\newfloat{graph}{htbp}{loh}

\usepackage{chemformula} 
\usepackage[version = 4]{mhchem} 

\usepackage{enumitem}
\usepackage{siunitx}
\usepackage[acronym]{glossaries}

\newacronym{aaa}{AAA}{adaptive Antoulas--Anderson}
\newacronym{apd}{APD}{avalanche photo-diode}
\newacronym{cbg}{CBG}{circular Bragg grating}
\newacronym{dbr}{DBR}{distributed Bragg reflector}
\newacronym{dbt}{DBT}{dibenzoterrylene}
\newacronym{dos}{DOS}{density of states}
\newacronym{fem}{FEM}{finite-element method}
\newacronym{fib}{FIB}{focused ion-beam}
\newacronym{fsr}{FSR}{free spectral range}
\newacronym{hbt}{HBT}{Hanbury Brown and Twiss}
\newacronym{hom}{HOM}{Hong-Ou-Mandel}
\newacronym{na}{NA}{numerical aperture}
\newacronym{pdcb}{\textit{p}-DCB}{para-dichlorobenzene}
\newacronym{pmma}{PMMA}{poly(methyl methacrylate)}
\newacronym{pva}{PVA}{polyvinyl alcohol}
\newacronym{qd}{QD}{quantum dot}
\newacronym{qkd}{QKD}{quantum key distribution}
\newacronym{smf}{SMF}{single-mode fiber}
\newacronym{sps}{SPS}{single-photon source}
\newacronym{bo}{BO}{Bayesian optimization}
\newacronym{gp}{GP}{Gaussian process}
\DeclareUnicodeCharacter{0229}{\c{e}}
\DeclareUnicodeCharacter{0228}{\c{E}}

\usepackage{authblk}
\author[1, 2]{Tim Hebenstreit}
\affil[1]{Max Planck Institute for the Science of Light, 91058 Erlangen, Germany}
\affil[2]{Department of Physics, Friedrich Alexander University Erlangen-Nuremberg,
91058 Erlangen, Germany}

\author[1, 2]{Jaime Gimeno Balaguer}

\author[3]{Fridtjof Betz}
\affil[3]{Zuse Institute Berlin, 14195 Berlin, Germany} 

\author[3]{Felix Binkowski}

\author[4, 5]{Maja Colautti}
\affil[4]{National Institute of Optics (CNR-INO), Sesto Fiorentino 50019, Italy} 
\affil[5]{European Laboratory for Non-Linear Spectroscopy (LENS), Sesto Fiorentino 50019, Italy} 

\author[1]{Jan Renger}

\author[4, 5]{Costanza Toninelli}

\author[3, 6]{Sven Burger}
\affil[6]{JCMwave GmbH, 14050 Berlin, Germany} 

\author[1, 2, 7]{Stephan Götzinger}
\affil[7]{Graduate School in Advanced Optical Technologies (SAOT), Friedrich Alexander University Erlangen-Nuremberg, 91052 Erlangen, Germany}

\newcommand{\SIauthors}{%
  Tim Hebenstreit$^{1,2}$,
  Jaime Gimeno Balaguer$^{1,2}$,
  Fridtjof Betz$^{3}$,
  Felix Binkowski$^{3}$,
  Maja Colautti$^{4,5}$,
  Jan Renger$^{1}$,
  Costanza Toninelli$^{4,5}$,
  Sven Burger$^{3,6}$,
  Stephan Götzinger$^{1,2,7}$%
}

\newcommand{\mytitle}{Design of monolithic microcavities \\ for enhancing organic quantum emitters}
\title{\mytitle}

\makeatletter
\newcommand{\savedauthors}{}
\newcommand{\savedaffils}{}
\AtBeginDocument{%
  \let\savedauthors\AB@authors
  \let\savedaffils\AB@affillist
}
\makeatother

\date{*Email: stephan.goetzinger@mpl.mpg.de}

\usepackage{comment}
\usepackage{amssymb}
\usepackage{hyperref}
\usepackage{etoc} 

\begin{document}

\begin{refsection}


\twocolumn[ 

\begin{@twocolumnfalse} 

\maketitle 

\begin{abstract}
    \noindent
    Single organic molecules are a well-established platform for high-quality single-photon generation: they can emit lifetime-limited photons with high purity and indistinguishability. However, their emission is accompanied by a pronounced red-shifted phonon sideband and higher-order vibrational peaks, which reduce the fraction of photons emitted into the desired narrowband zero-phonon line. The standard approach to suppressing this unwanted emission is to integrate the emitter into a monolithic photonic nanostructure that provides Purcell enhancement. However, incorporating organic materials using clean-room techniques has proven challenging, and has in fact so far prevented their integration into monolithic microcavities altogether. As a result, efficient, narrowband organic single-photon sources for applications in quantum information processing have remained elusive despite their considerable promise. Here, we propose three monolithic microcavity designs tailored to provide sufficient Purcell enhancement for organic quantum emitters. The Purcell effect induced by these cavities preferentially enhances emission into the 0-0 zero-phonon line, increasing spectral purity, photon extraction, and shortening the excited-state lifetime, which in turn relaxes the requirements for generating indistinguishable photons. The cavities have been optimized using Bayesian optimization and the \acrfull{aaa} algorithm for rational approximation, which offer global optimization and the efficient reconstruction of spectra from scattering simulations, respectively. These structures are compatible with both standard clean-room processing and single-molecule preparation techniques. Therefore, they offer a clear route to realize high-quality, practically monochromatic organic single-photon light sources.
\end{abstract}

\vspace{1cm}

\end{@twocolumnfalse} ]

\noindent
Single organic molecules have long served as a versatile platform for investigating light-matter interaction at the single-emitter level \cite{Moerner1989OpticalSolid, Lounis2000SingleTemperature}. Beyond their role as model systems, single molecules are also attractive for quantum technologies because they can operate as triggered single-photon sources with high purity and high photon indistinguishability \cite{Kiraz2005IndistinguishableMolecule, Rezai2018CoherenceNetworks, Lombardi2021TriggeredMolecule}. Among the various molecular emitters, \acrfull{dbt} embedded in crystalline organic hosts has emerged as one of the most widely used and best-performing systems, with demonstrations of high single-photon purity and two-photon interference from remote molecules \cite{Lombardi2021TriggeredMolecule, Lettow2010QuantumMolecules, Duquennoy2022Real-timeChip, Huang2025On-chipMolecules}. 

DBT in well-ordered matrices typically exhibits a comparatively large branching ratio of more than $30\,\%$, so that a substantial fraction of the emission can occur on the lifetime-limited 0-0 zero-phonon line (00ZPL) \cite{Verhart2016SpectroscopyPara-Dichlorobenzene, Musavinezhad2023QuantumTemperatures, Shkarin2026OrganicSources}. This branching ratio is sufficient for many proof-of-principle demonstrations but ultimately limits the application of organic emitters whenever the efficiency of indistinguishable photon generation matters, since part of the fluorescence is emitted into red-shifted vibronic transition lines rather than the 00ZPL. In addition, only a limited fraction of the emitted light can be collected in a single well-defined optical mode with standard free-space optics. Optical microcavities offer a direct route to address both issues: by enhancing the optical density of states at the 00ZPL frequency (Purcell effect)~\cite{Purcell1946SpontaneousFrequencies}, a resonant cavity can increase the emission rate in the 00ZPL relative to other decay channels and simultaneously funnel emission into a single cavity mode with a controlled spatial profile, improving collection in a Gaussian-type fundamental waveguide mode \cite{Lodahl2015InterfacingNanostructures, Wang2019TurningSystem}. The associated reduction of the excited-state lifetime not only enables operation at higher photon emission (repetition) rates but also relaxes the conditions for attaining indistinguishable photons.

Cavities for molecular emitters can be broadly divided into tunable open-access resonators and fully integrated (monolithic) microcavities. Throughout this work, monolithic refers to a mechanically rigid device without movable parts, whose resonance is fixed at fabrication and which therefore requires no active stabilization. Open cavities are attractive because they provide in situ tuning of the cavity resonance to the molecular 00ZPL and can produce a strong field enhancement at the emitter position \cite{Toninelli2010AEmission, Wang2017CoherentMicrocavity}. However, open cavities generally require picometer-scale control of the distance between mechanically movable mirrors and are therefore susceptible to acoustic noise and drift, often necessitating active stabilization to maintain alignment and resonance frequencies over long measurement times \cite{Wang2017CoherentMicrocavity, Wang2019TurningSystem}. Monolithic cavities, by contrast, promise intrinsic mechanical stability and a more straightforward path to scalable integration with on-chip photonic circuitry \cite{Lodahl2015InterfacingNanostructures, OBrien2009PhotonicTechnologies, Toninelli2021SingleTechnologies}.

For organic emitters, the main difficulty of integration into a monolithic structure is technological: many organic host crystals and molecular films are not compatible with standard nanofabrication steps (e.g. resist processing, plasma etching, high temperatures, vacuum conditions, and aggressive solvents). So far it has only been shown that it is possible to structure a polymer containing organic nanocrystals \cite{Ciancico2019NarrowPolymer}. The monolithic integration of molecules into a microcavity that modifies the spontaneous emission rate has not yet been demonstrated. In principle, the fabrication of photonic structures can be separated by exploiting the evanescent coupling of single molecules to waveguides and microresonators \cite{Tuerschmann2017, Lombardi2018PhotostableLight, Boissier2020CouplingMicro-structures, Shkarin2021NanoscopicMolecules, Rattenbacher2023On-chipMolecules, Huang2025On-chipMolecules, Lange2026AInteractions}. However, this has the drawback that crystalline matrices around these structures reduce the $Q$ factor of a cavity and that molecules close to the organic-inorganic interface, which are the ones that couple efficiently, exhibit compromised optical properties.

An advantage of molecular quantum emitters is their size: being only about a nanometer across, they can be integrated with high density. Thus, even if the mode volume of a cavity is small (such as in a \acrfull{cbg} cavity), there will be molecules in the correct spatial position. Together with an inhomogeneous broadening on the order of a few nanometers \cite{Nicolet2007SinglePhotophysics, Musavinezhad2024High-ResolutionCrystals}, one can thus find molecules whose 00ZPL frequency lies close to a targeted cavity resonance more easily \cite{Rickert2025AClock-rates}.

In this work, we optimize monolithic cavity designs for cryogenic applications specifically for DBT while keeping the host material flexible: anthracene (Ac) or \acrfull{pdcb}, depending on the molecule preparation and integration route \cite{Faez2014CoherentNanoguide, Pazzagli2018Self-AssembledEmission, Musavinezhad2024High-ResolutionCrystals}. DBT and these two hosts serve as representative examples; the designs apply equally to other fluorescent molecules and crystalline hosts such as para-terphenyl or naphthalene because they depend on the host primarily through its refractive index and the emitter's photophysical parameters. We present numerical design optimizations of three different microcavity designs, each based on a different fabrication approach. The central design goal is to realize cavities that (1) direct more than $80\,\%$ of the emitted power into a well-defined output mode and (2) ensure that more than $90\,\%$ of the collected photons originate from the molecular 00ZPL, while remaining compatible with realistic preparation and integration workflows.

\section{Design approach and considerations}

For molecules in crystalline hosts, the aforementioned target directly translates into engineering both the \textit{spectral} redistribution of spontaneous emission (enhanced 00ZPL emission) and the \textit{spatial} redistribution (efficient funneling into a single collectible optical mode). We note that throughout this work, $F_P$ denotes the multiplicative Purcell enhancement of the radiative rate, with $F_P=1$ in the absence of a cavity and $F_P = P_\mathrm{tot}/P_\mathrm{b}$ as obtained directly from our simulations (see Supporting Information Sec.~\ref{sec:derivations} for more details).

The spectral redistribution is captured by the fraction of photons emitted into the 00ZPL in the presence of a Purcell enhancement $F_P$, which can be regarded as a cavity-modified effective Franck-Condon factor $\alpha'$ (for derivation, see Supporting Information Sec.~\ref{sec:derivations}):
\begin{equation}
    \alpha' = \frac{F_P \cdot \alpha_\text{FC}}{\alpha_\text{FC} \left( F_P - 1 \right) + \frac{1}{\mathrm{QE}}},
    \label{eq:alpha_prime}
\end{equation}
where $\alpha_\mathrm{FC}$ is the free-space Franck-Condon branching into the 00ZPL and $\mathrm{QE}$ accounts for a possible non-ideal internal quantum efficiency. This expression emphasizes that significant improvements in 00ZPL performance can already be obtained at moderate $F_P$, because the cavity efficiently counteracts imperfect $\mathrm{QE}$ and limited $\alpha_\mathrm{FC}$ by preferentially enhancing the targeted radiative channel. For example, a Purcell enhancement of $F_P = 20$ leads, for a typical $\alpha_\text{FC} \approx 0.33$ and a $\mathrm{QE}$ of $85\,\%$, to an emission of about $90\,\%$ of the photons via the desired 00ZPL.

The spatial redistribution is quantified by the spontaneous emission coupling factor $\beta$, i.e. the fraction of the total spontaneous emission that is funneled into the cavity mode. In the convention used here, the rate coupled into the mode is the \emph{added} desired rate $\propto \alpha_\text{FC}(F_P - 1)$, so that (for derivation, see Supporting Information Sec.~\ref{sec:derivations}):
\begin{equation}
    \beta
    = \frac{1}{1 + \left[ \mathrm{QE}\cdot\alpha_\text{FC} \left( F_P - 1 \right) \right]^{-1}}.
    \label{eq:beta}
\end{equation}
The product of the two factors,
\begin{equation}
    \eta_\text{ZPL} = \alpha' \cdot \beta,
    \label{eq:eta_zpl}
\end{equation}
gives the overall probability that a single excitation is emitted into the 00ZPL \emph{and} collected through the cavity mode, i.e. the expected fraction of desired photons delivered into the output mode. Equation~\ref{eq:eta_zpl} is useful as a design target because it depends only on the emitter parameters and $F_P$, and can therefore be evaluated before any specific geometry is simulated. For the optimized structures below, $\beta$ is superseded by the directly simulated collection efficiency $\eta$ (Eq.~\ref{eq:eta}), which quantifies the same spatial redistribution without relying on the assumption that the entire added rate reaches the collection optics.

The considerations above require single molecules to be positioned at the electric field maximum of the cavity, embedded in a high-quality organic crystal. To this end, several strategies for the positioning and integration of DBT into crystalline hosts can be employed. One approach is to fill nano-channels or nano-capillaries with DBT-doped \acrshort{pdcb} \cite{Faez2014CoherentNanoguide, Maser2016Few-photonMolecule, Zirkelbach2022High-resolutionCrystal}. The doped \acrshort{pdcb} is heated to $T \approx \SI{60}{\celsius}$ and introduced in the molten state into channels that are narrow in height but extended laterally (typically $\approx \SI{400}{\nano\metre}$ height and $\approx \SI{500}{\micro\metre}$ width), the height being the parameter that sets the cavity length, where it solidifies into a crystalline structure. A second method is drop-casting DBT:Ac nanocrystals \cite{Pazzagli2018Self-AssembledEmission}, where nanocrystals prepared in solution are deposited onto a substrate and protected (e.g., using a \acrfull{pva} coating) to reduce degradation and evaporation effects. More recently, nanoprinting based on electrohydrodynamic dripping has enabled the controlled placement of organic crystalline material with high spatial resolution \cite{Musavinezhad2024High-ResolutionCrystals}. Finally, large sublimated crystals with optical-grade surfaces can also be employed \cite{Wei2020Single-molecule-dopedDevices, Keni2026VaporCircuits}. These preparation routes directly constrain viable cavity architectures, in particular with respect to material stacks, accessible thicknesses, and the need for fabrication steps that avoid resists, plasma processing, or harsh solvents in contact with the organic host.

Guided by these constraints and goals, we evaluate candidate monolithic microcavity designs using \acrfull{fem} simulations of the time-harmonic Maxwell's equations. The simulations provide access to the cavity mode profiles, field confinement at the emitter position, and Purcell enhancement, allowing us to quantify the expected improvement in 00ZPL fraction and collection efficiency under realistic assumptions for DBT in organic crystals. The numerical method and the extraction of the relevant figures of merit are described in the following section.

\section{Numerical Optimization}

The numerical simulations are performed with the \acrshort{fem}-solver \textsc{JCMsuite}~\cite{Burger2008JCMsuite:Nano-Optics}. All considered designs possess a cylindrical symmetry, which allows us to perform simulations on a 2D mesh with the source and the solution field being decomposed into a Fourier series in the azimuthal angle $\varphi$. Although the expansion is based on a cylindrically symmetric geometry, the source and solution fields are not required to be symmetric~\cite{Schneider2018NumericalFiber}, thus enabling the simulation of horizontal dipoles. The source is modeled as a transverse-electric polarized point source located on the symmetry axis. The total emitted power is given by $P_\mathrm{tot}=-0.5\,\mathrm{Re}(\mathbf{E}(\mathbf{r}_0,\omega)\cdot\mathbf{j}^*)$, where the electric field strength $\mathbf{E}$ is evaluated at the dipole position $\mathbf{r}_0$, $\omega$ is the excitation frequency, and $\mathbf{j} = -i\omega\mathbf{p}$ is the vector strength of the dipole with dipole moment $\mathbf{p}$. Normalization with the power emitted in homogeneous bulk material $P_\mathrm{b}$ yields the first figure of merit, which is the Purcell enhancement
\begin{equation}
    F_P = \frac{P_\mathrm{tot}}{P_\mathrm{b}}.
    \label{eq:F_P}
\end{equation}
The second figure of merit is the collection efficiency 
\begin{equation}
    \eta = \frac{P_\mathrm{NA}}{P_\mathrm{tot}},
    \label{eq:eta}
\end{equation}
where $P_\mathrm{NA}$ is the power emitted into the given \acrfull{na}. The aim is to achieve maximum collection efficiency under the constraint of $F_P>20$, which is sufficient for a fraction of about $90\,\%$ of the photons in the 00ZPL, as has been pointed out in the previous section. To achieve this goal, we use \acrfull{bo}. In the course of the optimization, a \acrfull{gp} is trained as a stochastic surrogate model of the target function~\cite{Garcia-Santiago2021BayesianStructures}. The target function is the maximum collection efficiency within a wavelength range centered at the target wavelength. In order to force the optimum to the center, we use a Gaussian envelope. The spectrum at a given parameter set suggested by \acrshort{bo} is efficiently evaluated using rational approximation with the \acrshort{aaa} algorithm~\cite{Nakatsukasa2018TheApproximation,Betz2024EfficientAlgorithm}. The optimization strategy begins with a reduction of the parameter space guided by theoretical considerations, followed by \acrshort{bo}. Finally, parameter scans are performed in the vicinity of the optima in order to quantify the robustness.

\begin{figure*}[t]
  \centering
  \includegraphics[width=5.75in]{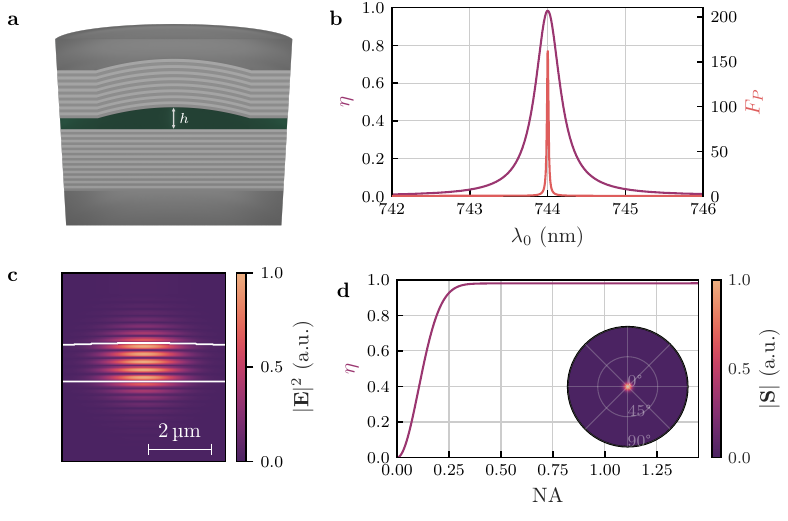}
  \caption{\textbf{Monolithic DBR-based Fabry--Perot cavity for DBT:\acrshort{pdcb}.} \textbf{a} Schematic cross-section with planar bottom DBR and concave top DBR ($R=\SI{50}{\micro\metre}$). \textbf{b} Simulated resonant Purcell enhancement ($F_P\approx 160$, $\approx\SI{0.02}{\nano\metre}$ bandwidth) and collection $\eta \approx 98\,\%$ into $\mathrm{NA} = 0.75$. \textbf{c} Normalized field intensity showing $h\approx \tfrac{5}{2}\lambda_0/n$, $V_\mathrm{eff} \approx 16 (\lambda_0/n)^3$, and $Q \approx 34000$. \textbf{d} Collection vs. numerical aperture, with emission mainly within $\mathrm{NA} \lesssim 0.3$. Inset: far-field radiation profile of the structure.}
  \label{fig:Fabry-Perot}
\end{figure*}

\section{Fabry--Perot cavity}

As a first design, we consider a Fabry--Perot microcavity aimed at simultaneously providing strong Purcell enhancement and high collection efficiency for DBT emission into the fundamental, approximately Gaussian, cavity mode. The target operating regime is a moderate mode volume ($V_\mathrm{eff} \approx 5$--$50 (\lambda_0/n)^3$) combined with a high quality factor ($Q \approx 10^4$), which is advantageous for efficient out-coupling while still allowing substantial enhancement of the emission rate.

Fabry--Perot cavities of this type have previously been realized mainly as open-access systems, where a planar and a curved mirror are independently fabricated and mechanically positioned with respect to each other to tune the cavity length and resonance \cite{Kelkar2015SensingVolume, Wang2017CoherentMicrocavity, Wang2019TurningSystem, Pscherer2021Single-MoleculeLevel, Tomm2021APhotons}. Here, we translate the same optical concept into a monolithic implementation using two dielectric \acrfull{dbr} mirrors: a planar bottom \acrshort{dbr} and a concave top \acrshort{dbr}. While the final performance is evaluated via FEM simulations, Gaussian beam optics provides a useful analytical design tool to pre-select a realistic parameter space -- most importantly the radius of curvature of the concave mirror -- thereby reducing the computational optimization effort.

A first-order analytical treatment of a planar-concave microcavity using Gaussian beam optics (see \cite{Greuter2014ACharacterization} and Supporting Information Sec.~\ref{sec:SI_FP}) with top mirror radius $R=\SI{50}{\micro\metre}$, cavity height $h \approx \tfrac{5}{2}\lambda_0/n$, cavity refractive index $n=1.6$ and theoretical top and bottom reflectivities $\mathcal{R}_1, \mathcal{R}_2 = 0.998, 0.99998$ yields a beam waist of $w_0 \approx \SI{1.06}{\micro\metre}$, a quality factor $Q \approx 1.5 \cdot 10^4$ (Finesse $\mathcal{F} \approx 3.1 \cdot 10^3$), an effective mode volume $V_{\mathrm{eff}} \approx 10 (\lambda_0/n)^3$ and emission into a numerical aperture of $\text{NA} \approx 0.22$. This analytical estimate serves only to pre-select the parameter space; the assumed mirror reflectivities are lower than those of the final quarter-wave stacks, which is why the FEM simulation below yields a quality factor about twice as large. The choice of $R$ presents a trade-off: a smaller $R$ reduces the mode volume and increases the Purcell enhancement but leads to a larger divergence angle, demanding a higher-NA objective for photon collection. 

With the Fabry--Perot cavity geometry established, the structure was parametrized, as shown in Fig.~\ref{fig:mesh}, and simulated using the method explained in the previous section. Table~\ref{tab:Fabry-Perot_parameters} lists the parameters used in the simulation. The cavity height $h$ was treated as a free parameter to find an enhanced resonance in a fundamental mode.

The optimized design (Fig.~\ref{fig:Fabry-Perot}) uses a concave top mirror made of 8 bilayers \ce{SiO2} and \ce{TiO2} with radius $R=\SI{50}{\micro\metre}$ and a bottom mirror made of 12 bilayers of the same materials. The structure supports a narrowband resonant enhancement of $F_P \approx 160$ with a linewidth of $\approx \SI{0.02}{\nano\metre}$ (Fig.~\ref{fig:Fabry-Perot}b). The simulated field intensity distribution along the cavity axis shows five intensity maxima, corresponding to a physical cavity length $h \approx \tfrac{5}{2}\lambda_0/n$, which produces an effective mode volume on the order of $V_\mathrm{eff} \approx 16 (\lambda_0/n)^3$ and a quality factor $Q \approx 34000$ (Fig.~\ref{fig:Fabry-Perot}c). In addition to the enhancement at the design frequency that favors 00ZPL emission, the cavity provides excellent far-field directionality: the simulated free-space collection efficiency reaches $\eta \approx 98\,\%$ in a numerical aperture of $\text{NA} = 0.75$ (Fig.~\ref{fig:Fabry-Perot}b), with most of the radiated power concentrated within $\mathrm{NA} \lesssim 0.3$ (Fig.~\ref{fig:Fabry-Perot}d), indicating good prospects for coupling to a single-mode optical fiber. As shown in Fig.~\ref{fig:fp_optimization}, the resonance can be shifted by $\pm\SI{1}{\nano\metre}$ while maintaining nearly the same peak Purcell enhancement. This corresponds to a cavity-length change of approximately $\pm\SI{2.1}{\nano\metre}$, which can be accommodated by the tapered trench geometry described below.

The concave top mirror substrate can be fabricated by \acrfull{fib} milling, grayscale lithography followed by etching, or laser ablation \cite{Wang2019TurningSystem, Hunger2010AFinesse}. The bottom substrate is etched to incorporate a trench that serves as a filling channel for DBT-doped \acrshort{pdcb} (as described in the integration approaches above). The trench is tapered in depth to ensure that, within some region, the correct cavity length for the molecular transition is achieved. This geometry also naturally accommodates, if needed, the integration of electrodes -- such as interdigitated ITO -- to enable Stark-effect tuning of molecular resonances to the cavity mode.

After substrate patterning, both mirrors are realized as DBRs consisting of alternating quarter-wave layers of $\ce{TiO2}$ and $\ce{SiO2}$ designed for $\lambda_0=\SI{744}{\nano\metre}$, corresponding to the 00ZPL of DBT:\acrshort{pdcb} \cite{Verhart2016SpectroscopyPara-Dichlorobenzene}. When realized experimentally, the DBR stop band should be chosen such that the 00ZPL lies within the high-reflectivity region while the excitation wavelength near $\SI{728}{\nano\metre}$ (0-1 excitation) is transmitted \cite{Nobakht2025HybridizationMode, Shkarin2026OrganicSources}. To enforce directional emission, the bottom DBR contains more layer pairs than the top DBR, so that photons are preferentially out-coupled upward. The multilayers can be deposited by ion beam sputtering (IBS) or alternatively by atomic layer deposition (ALD). Finally, the processed top and bottom substrates are bonded to form the monolithic cavity, and the organic crystal is introduced by channel filling to produce the complete device (Fig.~\ref{fig:Fabry-Perot}a) \cite{Gmeiner2016SpectroscopyDepth}.

Simulations of the dipole-position dependence allow the molecules to sit within a circle of radius $r\approx\SI{1.1}{\micro\metre}$ while maintaining $F_P > 20$. Along the optical axis, the standing-wave profile means that approximately $75\,\%$ of positions within one period still satisfy $F_P > 20$. This cavity design is therefore rather forgiving with respect to the positioning of the molecule.

\begin{figure*}
  \centering
  \includegraphics[width=5.5in]{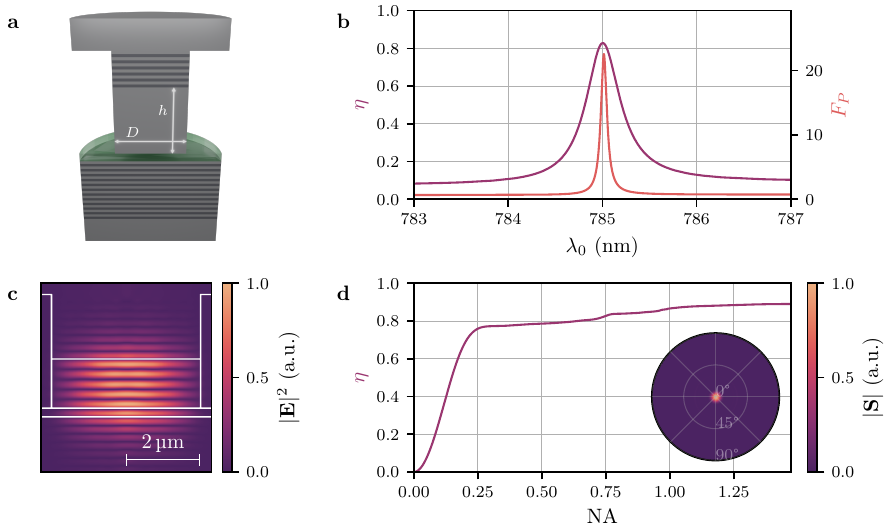}
  \caption{\textbf{DBR-based micropillar cavity for DBT:Ac nanocrystals.} \textbf{a} Cross-section and design parameters of the structure. \textbf{b} Simulated Purcell enhancement and collection efficiency into $ \mathrm{NA}=0.75$. \textbf{c} Normalized field intensity of the dominant eigenmode with $V_\text{eff} \approx 34(\lambda_0/n)^3$ and $Q \approx 10200$. \textbf{d} The collection efficiency as a function of the numerical aperture. The inset shows the energy radiated to the far-field. }
  \label{fig:micropillar}
\end{figure*}

\section{Micropillar cavity}

An evolution of the Fabry--Perot design is the micropillar cavity, where light is confined vertically by Bragg reflection from two DBR mirrors -- typically with asymmetric reflectivities to favor outcoupling through the top -- and laterally by the refractive index contrast at the etched sidewalls. This structure derives from a simple planar microcavity: the lateral patterning turns the in-plane continuum into discrete modes with a well-defined mode volume, and thus enables a significant Purcell enhancement. Such structures have been widely investigated in the context of semiconductor quantum dots, which yield high extraction efficiency (typically exceeding $70\,\%$) and moderate Purcell enhancement ($5$ to $15$ in state-of-the-art devices) \cite{Lodahl2015InterfacingNanostructures, Somaschi2016Near-optimalState, Unsleber2016HighlyEfficiency, Gines2022HighCavity}. The emission enhancement in micropillars can be interpreted as the combined effect of Fabry--Perot cavities and dielectric antennas. Because the solid angle subtended by the cavity mode is not negligible compared to $4\pi$, the emission rate into non-cavity (side) modes, $\Gamma_\mathrm{side}$, can no longer be equated with the free-space rate $\Gamma_0$. Therefore, the geometry is chosen to suppress side losses, allowing the same Purcell enhancement to be obtained at lower quality factors~\cite{Dlaka2024DesignCavities, Jordan2025TheMicrocavity}. In particular, Fig.~\ref{fig:pillar_analysis}d of the Supporting Information shows that the collection efficiency exhibits modulations as a function of the pillar diameter, suggesting that leaky-mode suppression contributes to determining the device performance \cite{Dlaka2024DesignCavities, Jordan2025TheMicrocavity}.

In the proposed design, depicted in Fig.~\ref{fig:micropillar}a, we consider a micropillar cavity made of glass with DBR mirrors at the top and bottom facets. The monolithic structure integrates a sublimated anthracene crystal doped with DBT molecules ($\lambda_0=\SI{785}{\nano\metre}$) at its base. In order to maximize extraction, the top DBR comprises 8 $\lambda/4$-thick \ce{SiO2}/\ce{TiO2} bilayers, the bottom DBR 12. The optimized design shown in Fig.~\ref{fig:micropillar}b corresponds to a pillar diameter of $D = \SI{4040}{\nano \metre}$ and a glass height of $h = \SI{1332}{\nano \metre}$, to which the \SI{240}{\nano\metre} crystal adds and yields $F_P \approx 23$, $Q \approx 10200$, $V_\text{eff} \approx 34(\lambda_0/n)^3$, and $\eta \approx 83\,\%$ into $\mathrm{NA}=0.75$. Interestingly, the extraction efficiency approaches values $>80\,\%$ already for $\text{NA}<0.5$.

The geometrical configuration is mostly parametrized by the pillar aspect ratio $D/h$, defining the mode characteristics with a pronounced antinode at the DBT emitter position, thereby maximizing the coupling efficiency. The pillar height is finally chosen in order to spectrally match the central wavelength of the DBT:Ac inhomogeneous broadening.

With narrower pillars, the quality factor decreases for comparable Purcell enhancement, whereas the far-field pattern becomes more structured, with correspondingly lower collection efficiency (see Fig.~\ref{fig:pillar_analysis}). This option can nevertheless be attractive because it reduces the sensitivity to fabrication imperfections and relaxes the constraints on the tuning of emitters, which would otherwise require additional control mechanisms and add technical complexity.

From a fabrication point of view, the proposed micropillar devices can be realized using standard nanofabrication techniques. The top DBR and the glass layer are patterned into micropillar structures by high-resolution electron-beam lithography and reactive ion etching (RIE). In a second step, molecular crystals are positioned on the extended bottom DBR mirror, and the fabricated pillars are aligned onto them via simple translational control, thus avoiding any detrimental processing of the emitters. We note that this geometry constrains the emitter to the bottom plane of the cavity, which is compatible with large-area sublimated crystals extending over tens to hundreds of micrometers laterally while remaining only a few hundred nanometers thick. A single crystal can therefore serve an entire array of micropillars, increasing the likelihood that one or more molecules are naturally positioned within the optical mode of a given cavity. This relaxes the need for deterministic placement while supporting parallel device integration.

Importantly, the proposed architecture is not limited to this etched-glass implementation, highlighting its fabrication flexibility. As an alternative, the glass pillar can be replaced by a polymeric photoresist, written either directly onto the pre-structured top DBR or on the molecular crystal via two-photon polymerization~\cite{Maruo1997Three-dimensionalPhotopolymerization, Wu2019Polymer-BasedTechnology}. This approach has proven crucial for the realization of increasingly complex photonic components such as tritters or fiber-to-chip couplers~\cite{Spagnolo2013Three-photonTritter, Schumann2014HybridWriting}. When the emitters are exposed to the resist and the writing beam, a thin protective layer is essential: it has been demonstrated that DBT molecules in anthracene crystals are compatible with the full lithographic process once properly embedded in such a layer~\cite{Colautti2020ATechnologies, Lombardi2024AdvancesMolecules}. A similar methodology has also been adopted for the coupling of epitaxially grown quantum dots to the waveguide mode of polymer nanowires, enhancing collection efficiency into a single-mode fiber~\cite{Perez2025Direct-Laser-WrittenMode}.

A third method separates the polymer pillar fabrication from the organic host and DBR, implementing direct laser writing on conformable freestanding membranes~\cite{MarcoDenHoed2023FacileMembranes}. The patterned film is then transferred onto the anthracene crystal and the structure is completed upon alignment of the structured top DBR. The conformal transfer avoids exposing the molecular crystal to the deposition process and could thus eliminate any fabrication-induced damage.

\section{Circular Bragg grating cavity}

\begin{figure*}
  \centering
  \includegraphics[width=6 in]{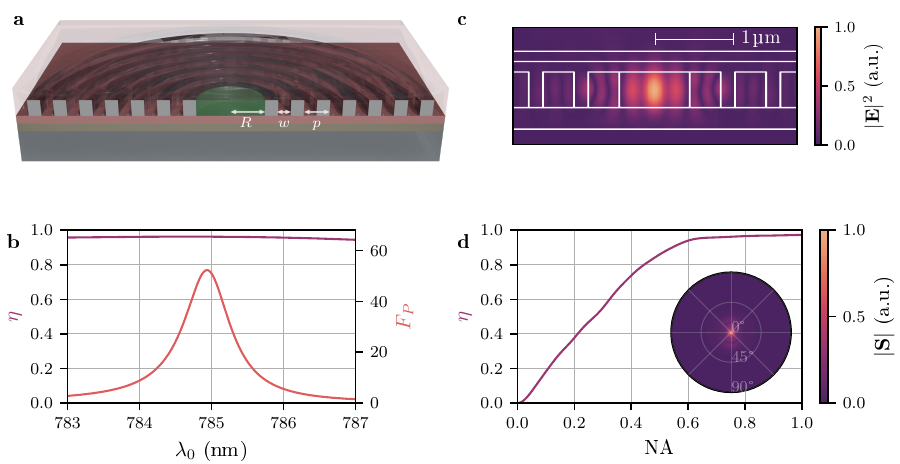}
  \caption{\textbf{Circular Bragg grating cavity for nanoprinted DBT:Ac.} \textbf{a} \acrshort{cbg} cross-section and design parameters. \textbf{b} Simulated Purcell enhancement $F_P \approx 52$ (bandwidth $\approx \SI{1}{\nano \metre}$) and free-space collection efficiency $\eta \approx 96\,\%$ into $\mathrm{NA}=0.75$. \textbf{c} Normalized field intensity with $V_\text{eff}\approx 1.3 \left(\frac{\lambda_0}{n}\right)^3$, and $Q \approx 900$, showing strong confinement within the central disk and first ring. \textbf{d} Collection vs. numerical aperture, with emission mainly within $\mathrm{NA} \lesssim 0.75$. Inset: far-field radiation profile of the structure. } 
  \label{fig:CBG}
\end{figure*}

Lastly, we consider a \acrshort{cbg} cavity. As shown in Fig.~\ref{fig:CBG}a, this design exploits cylindrical symmetry to form concentric rings that constitute a radial grating. The central disk consists of the DBT-doped anthracene crystal itself, surrounded by concentric \ce{TiO2} rings of the same height $h_\mathrm{disk}$, so that the emitter sits in a low-index defect at the center of a high-index-contrast radial grating. The dipole emitter is located at the center of the structure and emits primarily along the radial axis. Following the approach of reference \cite{Davanco2011AEmission} (and illustrated in Fig.~\ref{fig:cbg_coupler}), the grating period is optimized to out-couple the radial mode vertically, producing a far-field that resembles a Gaussian mode (see Fig.~\ref{fig:CBG_theory}c) and therefore matches a fundamental fiber mode for efficient coupling.

The cavity confines a Bessel-like leaky mode mainly within the central disk and the innermost rings (Fig.~\ref{fig:CBG_theory}b), which yields a small effective mode volume $V_{\mathrm{eff}}$. Since the grating is tuned closer to the vertical out-coupling condition than to the Bragg-reflection condition, the resulting $Q$ factor is lower than that of previously proposed resonators. However, the small $V_{\mathrm{eff}}$ compensates for this: a high Purcell enhancement remains achievable without requiring a high $Q$ (see Supporting Information Sec.~\ref{sec:SI_CBG}). In short, the \acrshort{cbg} confines the leaky radial mode while efficiently redirecting light vertically.

In the absence of a bottom mirror, the emission is radiated almost equally upward and downward. To maximize upward collection, a bottom silver reflector is added beneath the \acrshort{cbg}. A \ce{SiO2} spacer of thickness $\lambda_0/(2n)$ separates the silver from the active region to (1) prevent direct contact between the organic material and the metal and (2) ensure constructive interference of the reflected field in the upward direction. A detailed discussion of the choice of spacer thickness can be found in the Supporting Information (see Fig.~\ref{fig:CBG_parameter_sweeps}a).

The \acrshort{cbg} also offers practical advantages over the previous cavities. The lack of a top DBR and its stop band allows for an efficient optical excitation of the molecule. Fabrication is simplified by using nanoprinted molecules~\cite{Musavinezhad2024High-ResolutionCrystals}: DBT:Ac crystals, whose 00ZPL is centered at $\SI{785}{\nano\metre}$, can be placed in the central disk after patterning, avoiding complex deterministic alignment procedures \cite{Moczaa-Dusanowska2020Strain-TunableDots, Kolatschek2019DeterministicLithography}. However, the bandwidth of the Purcell enhancement is wider than for the two cavities with DBRs. This leads to the enhancement of a part of the phonon sideband accompanying the 00ZPL, which can negatively influence the photons' indistinguishability.

The optimization of the design depends sensitively on several geometric and material parameters: the radius of the central disk $R_{\mathrm{disk}}$, the grating period $p$, the ridge width $w$, and the layer heights $h_{\ce{Ag}}, h_{\ce{SiO2}}, h_{\mathrm{disk}}, h_{\mathrm{PMMA}}$ (Fig.~\ref{fig:CBG}a). To keep optimization computationally tractable, we selected the most influential free parameters: the radius of the central disk $R_{\mathrm{disk}}$ and the parameters that define the grating $w$ and $p$. The height parameters were fixed: $h_{\ce{SiO2}}=\lambda/2$ (as above), $h_{\mathrm{PMMA}} = h_{\mathrm{PVA}} =\lambda_0/(4n)$ (for constructive interference), and the \acrshort{cbg} disk height $h_{\mathrm{disk}}=\SI{450}{\nano\metre}$ -- shared by the anthracene disk and the \ce{TiO2} rings. This height is chosen from parametric sweeps (Fig.~\ref{fig:CBG_parameter_sweeps}b) as a practical optimum that balances optical performance, fabrication constraints, and numerical cost. Optimization confirmed that $R_{\mathrm{disk}}$ is a highly critical parameter: only narrow ranges of $R_{\mathrm{disk}}$ yield Purcell enhancement, and small changes in $R_{\mathrm{disk}}$ can be used to tune the resonance wavelength (Figs.~\ref{fig:CBG_theory}a, ~\ref{fig:CBG_parameter_sweeps}c) \cite{Rickert2019OptimizedGratings, Rickert2025AClock-rates}.

The optimized \acrshort{cbg} shown in Fig.~\ref{fig:CBG}b achieves a Purcell enhancement of $F_P\approx52$ on a narrow spectral band (bandwidth $\approx \SI{1}{\nano \metre}$). The mode is strongly confined within the central disk and the first ring (Fig.~\ref{fig:CBG}c), exhibiting Bessel-like oscillations consistent with reference \cite{Ates2012BrightMicrocavity} and the intuition gained in Supporting Information Sec.~\ref{sec:SI_CBG}. The effective mode volume is $V_\text{eff}\approx 1.3 \left(\frac{\lambda_0}{n}\right)^3$ and the quality factor is $Q\approx900$. The far-field (Fig.~\ref{fig:CBG}d) shows a near-Gaussian angular distribution with most power inside $\mathrm{NA} \lesssim 0.64$, corresponding to a free-space collection efficiency $\eta\approx96\,\%$ in an objective of $\mathrm{NA}=0.75$.

We further studied parameter dependencies: the number of grating rings strongly affects the out-coupling efficiency and narrowband enhancement (confirming the role of the grating; see Fig.~\ref{fig:CBG_parameter_sweeps}d), the dipole orientation in the anthracene matrix also influences the enhancement, following a cosine-like dependence (Fig.~\ref{fig:CBG_parameter_sweeps}e). Sensitivity to the position of the emitter was also investigated by varying both radial ($r$) and vertical ($z$) coordinates. As shown in Fig.~\ref{fig:CBG_parameter_sweeps}f, $F_P$ remains above $20$ for vertical displacements of $\pm\SI{175}{\nano\metre}$ and radial displacements of up to \SI{140}{\nano\metre}.

Lastly, a robustness analysis to maintain $\beta > 90\,\%$ over an inhomogeneous broadening of $\lambda_0 \pm \SI{1}{\nano\metre}$ yielded tolerances of $\pm\SI{2.9}{\nano\metre}$ for the radius of the disk, $\pm\SI{2.1}{\nano\metre}$ for the width of the grating, and $\pm\SI{1.5}{\nano\metre}$ for the period. While tight, these are theoretically within the placement accuracy of current electron-beam lithography systems, provided that resist and etching processes are carefully optimized.

We propose the following fabrication route. Starting from a \ce{SiO2} substrate, a thin ($\approx \SI{5}{\nano\metre}$) chromium adhesion layer is deposited to promote silver adhesion, followed by a $\SI{200}{\nano\metre}$ silver layer forming the back-reflector. A \ce{SiO2} spacer of thickness $\lambda_0/(2n)$ is then deposited to satisfy the constructive-reflection condition. Next, a \ce{TiO2} layer is deposited and patterned into concentric rings by electron-beam lithography and plasma etching. To account for fabrication tolerances, critical parameters will be systematically varied around their nominal values during the fabrication process. The structure is subsequently protected by a \acrshort{pmma} coating, after which the central disk is reopened via a second lithography step. DBT:Ac nanocrystals are then printed into the exposed central disk by electrohydrodynamic dripping~\cite{Musavinezhad2024High-ResolutionCrystals}. Finally, the completed device is covered with a thin \acrshort{pva} layer, which serves to (1) protect the structure and (2) prevent crystal evaporation. The resulting cross-section is illustrated in Fig.~\ref{fig:CBG}a.

\section{Discussion}

Performance metrics for the three proposed monolithic microcavity designs are summarized in Table~\ref{tab:summary}. Compared with bulk emitters, all three structures offer an improvement in both the spectral purity of the emission (via Purcell enhancement of the zero-phonon line) and the spatial collection efficiency. Crucially, all three designs meet the targets set at the outset of this work: the simulated collection efficiencies satisfy $\eta > 80\,\%$ in each case, and the Purcell factors achieved are well above the $F_P \approx 20$ shown to be sufficient for $> 90\,\%$ of collected photons to originate from the 00ZPL of DBT. Note that these designs are merely exemplary. For instance, adding more DBR pairs will in simulations always result in a larger theoretical Purcell factor.

The Fabry--Perot and micropillar cavities, relying on reflective DBR mirrors, achieve higher quality factors and narrower enhancement bandwidths than the \acrshort{cbg}, with performance levels comparable to those reported for state-of-the-art micropillar systems~\cite{Margaria2025EfficientPhotons}. However, narrow cavity linewidths come with a difficulty: matching the emitter frequency precisely to the cavity resonance. In particular, the resonant wavelength of the DBR-based cavities depends linearly on the cavity length (or pillar height), in principle requiring nanometer-scale precision to align the cavity mode with the 00ZPL of the target molecule. In practice, this constraint is relaxed by the nanometer size of molecular emitters: their high achievable density means that even a small mode volume will always contain molecules well positioned within the cavity mode. Precise spectral match is achieved through arrays of varying cavity length. Additional fine-tuning then relies on integrated electrodes (as discussed for the Fabry--Perot design) together with laser-induced frequency tuning \cite{Colautti2020Laser-inducedEmitters}.

In contrast, the \acrshort{cbg} cavity offers a more fabrication-friendly approach. It does not require complex multilayer DBR stacks, and its performance is reasonably tolerant to minor nanofabrication imperfections. The \acrshort{cbg} still achieves a high Purcell factor ($F_P \approx 52$) due to its small mode volume ($V_\text{eff}\approx 1.3 \left(\frac{\lambda_0}{n}\right)^3$). An additional practical advantage is its open top: with no top mirror in the excitation path, the molecule can be addressed directly from above by a focused laser beam, simplifying the experimental setup. However, the lower quality factor ($Q \approx 900$) results in a wider enhancement bandwidth. Although this is sufficient for high-brightness single-photon sources, it may limit the degree of photon indistinguishability due to the enhancement of the phonon sideband.

Beyond the optical performance, the three designs differ in their associated organic-integration routes -- channel filling of doped \acrshort{pdcb} for the Fabry--Perot, nanoprinting of DBT:Ac for the \acrshort{cbg}, and placement of sublimated crystals for the micropillar. This pairing of cavity architectures with distinct preparation methods is a deliberate feature: different laboratories can select the design best matched to their available fabrication and chemistry capabilities.
A critical remaining challenge for all three designs is the efficient interfacing of these sources with quantum networks. Although the simulated free-space collection efficiencies ($\eta > 80\,\%$) are promising, practical deployment requires coupling of this light into single-mode fibers or on-chip waveguides. The far-field profiles of these cavities — particularly the Gaussian-like emission of all three designs — suggest that high coupling efficiencies are achievable. Future work will focus on optimizing the optical train (e.g. using aspheric lenses and other linear optical components) to match the cavity output mode to the fundamental mode of a single-mode fiber, thereby minimizing coupling losses.

\begin{table}[ht!]
    \centering
    \small
    \setlength{\tabcolsep}{3pt} 
    \begin{tabular}{@{}l|ccc@{}}
        parameter & Fabry--Perot & micropillar & \acrshort{cbg} \\
        \hline
        wavelength $\lambda_0$ & $\SI{744}{\nano\metre}$ & $\SI{785}{\nano\metre}$ & $\SI{785}{\nano\metre}$ \\
        quality factor $Q$ & 34000 & 10200 & 900 \\
        mode volume $V_\text{eff}$ & $16 \left(\frac{\lambda_0}{n} \right)^3$ & $34 \left(\frac{\lambda_0}{n} \right)^3$ & $1.3 \left(\frac{\lambda_0}{n} \right)^3$ \\
        Purcell factor $F_P$ & 160 & 23 & 52 \\
        collection eff.\ $\eta$ & $98\,\%$ & $83\,\%$ & $96\,\%$\\
        bandwidth $\Delta\lambda$ & $\SI{0.02}{\nano\metre}$ & $\SI{0.08}{\nano\metre}$ & $\SI{0.9}{\nano\metre}$ \\ 
    \end{tabular}
    \caption{Summary of important parameters of all three cavities optimized in this work.}
    \label{tab:summary}
\end{table}

\section{Conclusion}

In summary, we have presented the design and numerical optimization of three monolithic microcavity architectures tailored for organic quantum emitters. Using the \acrshort{aaa} algorithm and Bayesian optimization, we identified designs that achieve a Purcell enhancement $F_P > 20$ and free-space collection efficiencies $\eta > 80\,\%$ for the 0-0 zero-phonon line of DBT molecules. To the best of our knowledge, these are the first proposed designs predicted to deliver Purcell-enhanced emission from single organic molecules integrated into monolithic microcavities, addressing a long-standing gap in the field. The Fabry--Perot and micropillar cavities are well-suited for the generation of indistinguishable single photons, whereas the \acrshort{cbg} provides a robust platform for high-brightness emission with simplified fabrication and direct optical access. Each architecture is paired with a distinct organic-integration route, offering flexibility across different laboratory capabilities. Crucially, the optimization framework employed here is generalizable; it can be readily adapted to tailor these cavity geometries for other emitter wavelengths, host materials, or specific fabrication constraints, with computational cost manageable on standard workstations. These designs contribute to the scalable fabrication of practical, narrowband organic single-photon sources for quantum technologies.

\section*{Acknowledgments}

T.H., J.G.B., J.R. and S.G. acknowledge financial support from the Deutsche Forschungsgemeinschaft (DFG, German Research Foundation) -- ID 429529648 -- TRR 306 QuCoLiMa (Quantum Cooperativity of Light and Matter), the Free State of Bavaria via the Munich Quantum Valley light house project "QuMeCo" and the Max Planck Society. F.Be., F.Bi. and S.B. acknowledge funding by the Deutsche Forschungsgemeinschaft (DFG, German Research Foundation) under Germany's Excellence Strategy -- The Berlin Mathematics Research Center MATH+ (EXC-2046/1, EXC-2046/2, project ID: 390685689) and by the German Federal Ministry of Research, Technology, and Space (BMFTR, Forschungscampus MODAL, project 05M20ZBM). This project (20FUN05 SEQUME) has received funding from the EMPIR programme co-financed by the participating states and from the European Union’s Horizon 2020 research and innovation programme. Additionally, M.C. and C.T. acknowledge funding by the European Union (ERC, QUINTESSEnCE, 101088394). Views and opinions expressed are however those of the authors only and do not necessarily reflect those of the European Union or the European Research Council. Neither the European Union nor the granting authority can be held responsible for them. T.H. is part of the Max Planck School of Photonics supported by the German Federal Ministry of Research, Technology, and Space (BMFTR), the Max Planck Society and the Fraunhofer Society. J.G.B. acknowledges financial support from the Erasmus+ Programme of the European Union, the Fundación Ramón Areces, Banco Santander, and the Generalitat Valenciana (GVA). The authors acknowledge the use of Claude Opus 4.8 to assist with language editing, grammar correction, and improving overall readability of the manuscript. The authors reviewed and edited all content generated by this tool and take full responsibility for the accuracy, data integrity, and final text of the published work.

\section*{Data availability}

The source code for the numerical simulations in this work can be found in the open access data publication: https://doi.org/10.5281/zenodo.20703172

\section*{Competing interests}
The authors declare no competing financial interest.

\section*{Supporting information}

Supporting Information available: Summary of optimal cavity parameters; derivation of the effective Franck-Condon factor (Eq.~\ref{eq:alpha_prime}) and the $\beta$ factor (Eq.~\ref{eq:beta}); further information on the Fabry--Perot, the micropillar, and the \acrshort{cbg} cavity.

\printbibliography[title={References}]
\end{refsection}

\clearpage

\section*{Table of contents graphic}

\includegraphics[width=3.25 in]{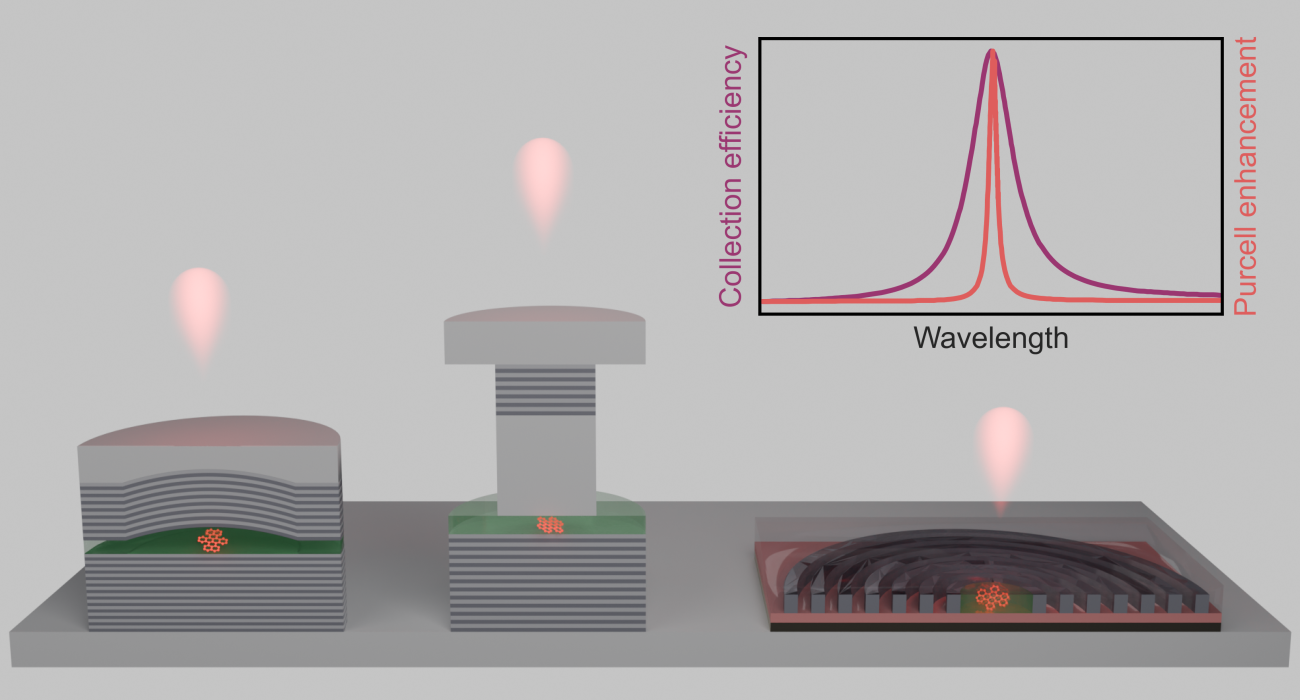}

\newpage

\clearpage
\onecolumn

\begin{center}
  {\LARGE Supporting Information for:}\\[0.5em]
  {\LARGE \mytitle}\\[1.5em]
  \SIauthors\\[1em]
  \savedaffils
\end{center}
\smallskip

\addcontentsline{toc}{part}{Supplementary Information}

\setcounter{page}{1}
\renewcommand{\thepage}{S\arabic{page}}

\setcounter{secnumdepth}{3}

\setcounter{section}{0}

\setcounter{figure}{0}
\setcounter{table}{0}
\setcounter{equation}{0}
\renewcommand{\thefigure}{S\arabic{figure}}
\renewcommand{\thetable}{S\arabic{table}}
\renewcommand{\theequation}{S\arabic{equation}}

\etocsettocdepth{section} 
\localtableofcontents


\begin{refsection}

\section{Summary of optimal cavity parameters}

In the following, the parameters that are important for the fabrication of each of the three cavities are given. Some of the parameters were chosen to be fixed in the simulation with \textsc{JCMwave} and some of them were variable. This is also indicated.

\subsection{Parameters for the Fabry--Perot cavity}

\begin{table}[ht!]
    \centering
    \begin{tabular}{ccc}
        parameter & value & variable / fixed \\
        \hline
        cavity height & \SI{1175.2}{\nano\metre} & variable \\
        radius of curvature of top DBR & \SI{50}{\micro\metre} & fixed \\
        layers of bottom DBR & 12 & fixed \\
        layers of top DBR & 8 & fixed \\
        \ce{TiO2} layer height of DBR & $\lambda_0/(4n)$ & fixed \\
        \ce{SiO2} layer height of DBR & $\lambda_0/(4n)$ & fixed \\
        refractive index of \acrshort{pdcb} & 1.6 & fixed \\
        refractive index of \ce{TiO2} & 2.42 & fixed \\
        refractive index of \ce{SiO2} & 1.45 & fixed \\
    \end{tabular}
    \caption{Parameters for the Fabry--Perot cavity. The refractive indices of \ce{SiO2} and \ce{TiO2} were measured using an ellipsometer. The layers were deposited using atomic layer deposition (ALD).}
    \label{tab:Fabry-Perot_parameters}
\end{table}

\subsection{Parameters for the micropillar cavity}

\begin{table}[ht!]
    \centering
    \begin{tabular}{ccc}
        parameter & value & variable / fixed \\
        \hline
        cavity diameter & \SI{4040}{\nano\metre} & variable \\
        cavity height & \SI{1332.3}{\nano\metre} & variable \\
        crystal height & \SI{240}{\nano\metre} & fixed \\
        layers of bottom DBR & 12 & fixed \\
        layers of top DBR & 8 & variable \\
        \ce{TiO2} layer height of DBR & $\lambda_0/(4n)$ & fixed \\
        \ce{SiO2} layer height of DBR & $\lambda_0/(4n)$ & fixed \\
        refractive index of \ce{Ac} & 1.7 & fixed \\
        refractive index of \ce{TiO2} & 2.27 & fixed \\
        refractive index of \ce{SiO2} & 1.47 & fixed \\
    \end{tabular}
    \caption{Parameters for the micropillar cavity.}
    \label{tab:micropillar_parameters}
\end{table}

\subsection{Parameters for the circular Bragg grating cavity}

\begin{table}[ht!]
    \centering
    \begin{tabular}{ccc}
        parameter & value & variable / fixed \\
        \hline
        inner disk radius & \SI{450}{\nano\metre} & variable \\
        period & \SI{570}{\nano\metre} & variable \\
        ridge width & \SI{178}{\nano\metre} & variable\\
        disk / ring height $h_\text{disk}$ & \SI{450}{\nano\metre} & fixed \\
        Ag height & \SI{200}{\nano\metre} & fixed\\
        SiO$_{2}$ height& \SI{270}{\nano\metre} &fixed \\
        PMMA height & $\lambda_0/(4n)$ & fixed \\
        \acrshort{pva} height & $\lambda_0/(4n)$ & fixed \\
        refractive index of \ce{Ac} & 1.7 & fixed \\
        refractive index of \ce{Ag} & 0.09+5.06i & fixed \\
        refractive index of \ce{TiO2} & 2.42 & fixed \\
        refractive index of \ce{SiO2} & 1.45 & fixed \\
        refractive index of \ce{PMMA} & 1.49 & fixed \\
        refractive index of \acrshort{pva} & 1.5 & fixed \\
        grating rings & 16 & fixed \\
    \end{tabular}
    \caption{Parameters for the circular Bragg grating cavity.}
    \label{tab:CBG_parameters}
\end{table}

\section{Derivations of Eq.~\ref{eq:alpha_prime} and Eq.~\ref{eq:beta} from main text}
\label{sec:derivations}

\subsection{Derivation of the effective Franck-Condon factor (Eq.~\ref{eq:alpha_prime}) from main text}
\label{subsec:Franck-Condon}
 
In the following derivation the Franck-Condon branching is $\alpha_{\text{FC}}$ and $\mathrm{QE}$ is the internal quantum efficiency. Let $\gamma_{\rm rad}$ be the free-space radiative rate and $\gamma_{\rm nr}$ the non-radiative rate. The free-space rates are therefore:
\begin{equation}
    \gamma_d = \alpha_{\text{FC}} \cdot \gamma_{\rm rad} \quad (\text{desired line}), 
    \qquad 
    \gamma_o = (1 - \alpha_{\text{FC}}) \cdot \gamma_{\rm rad} \quad (\text{other lines}).
\end{equation}
and from $\mathrm{QE} = \dfrac{\gamma_{\rm rad}}{\gamma_{\rm rad}+\gamma_{\rm nr}}$ one gets:
\begin{equation}
    \gamma_{\rm nr} = \gamma_{\rm rad} \left( \frac{1}{\mathrm{QE}} - 1 \right).
\end{equation}
Now, a cavity with Purcell factor $F_P$ is turned on. It is assumed that it only multiplies the desired radiative rate:
\begin{equation}
    \gamma'_d = F_P \cdot \gamma_d = F_P \cdot \alpha_{\text{FC}} \cdot \gamma_{\rm rad}, 
    \qquad 
    \gamma'_o = \gamma_o,
    \qquad 
    \gamma'_{\rm nr} = \gamma_{\rm nr}.
    \label{eq:radiative_rates}
\end{equation}
The total decay rate in a cavity is then:
\begin{equation}
    \Gamma_{\rm tot} = \gamma'_d + \gamma'_o + \gamma'_{\rm nr}
    = \gamma_{\rm rad} \left[ F_P \cdot \alpha_{\text{FC}} + (1 - \alpha_{\text{FC}}) + \left( \frac{1}{\mathrm{QE}} - 1 \right) \right]
    =\gamma_{\rm rad} \left[ \alpha_{\text{FC}} (F_P - 1) + \frac{1}{\mathrm{QE}} \right].
\end{equation}
The branching probability for the desired transition (i.e. the probability that one excitation yields a photon on that line) defines the cavity-modified, effective Franck-Condon factor:
\begin{equation}
    \alpha' = \frac{\gamma'_d}{\Gamma_{\rm tot}}
    = \frac{F_P \cdot \alpha_{\text{FC}} \cdot \gamma_{\rm rad}}
    {\gamma_{\rm rad} \left[ \alpha_{\text{FC}} (F_P - 1) + 1/\mathrm{QE} \right]}
    = \boxed{ \dfrac{F_P \cdot \alpha_{\text{FC}}}{\alpha_{\text{FC}}(F_P - 1) + \dfrac{1}{\mathrm{QE}}}}.
    \label{eq:alpha_prime_SI}
\end{equation}
The derivation assumes the weak-coupling (Purcell) regime, where the cavity modifies the spontaneous emission rate without coherent emitter-cavity dynamics. In this work, the Purcell factor $F_P$ is defined as the multiplicative enhancement of the radiative rate ($F_P = 1$ in the absence of a cavity) and only enhances the chosen spectral line. Other radiative and non-radiative rates are unchanged. Note that an additive convention is also common in the literature, where the cavity-modified rate is written as $(1+F)\gamma_0$ and $F=0$ corresponds to no cavity~\cite{Wang2019TurningSystem}. The two are related by $F_P=1+F$; the distinction is immaterial at the large values reported in this work, but is relevant when comparing Eq.~\ref{eq:alpha_prime} and Eq.~\ref{eq:beta} to expressions written in terms of $F$. The internal quantum efficiency $\mathrm{QE}$ is defined as above.
 
\subsection{Derivation of the $\beta$ factor (Eq.~\ref{eq:beta}) from main text}
 
The spontaneous-emission coupling factor $\beta$ is the fraction of the total emitted power that is funneled into the cavity mode. Using the same notation and rates as in the previous section, only the \emph{added} part of the desired radiative rate is coupled into the cavity mode. In the multiplicative convention used in this work, the free-space desired rate $\gamma_d$ is already present without the cavity, while the cavity adds $\gamma'_d - \gamma_d = (F_P - 1)\gamma_d = (F_P-1)\,\alpha_{\text{FC}}\,\gamma_{\rm rad}$. The $\beta$ factor is therefore the ratio of this cavity-coupled rate to the total decay rate $\Gamma_{\rm tot}$ derived above:
\begin{equation}
    \beta = \frac{\gamma_d (F_P - 1)}{\Gamma_{\rm tot}}
    = \frac{\alpha_{\text{FC}} \cdot \gamma_{\rm rad} \cdot (F_P - 1)}
    {\gamma_{\rm rad} \left[ \alpha_{\text{FC}} (F_P - 1) + 1/\mathrm{QE} \right]}.
\end{equation}
Cancelling $\gamma_{\rm rad}$ yields:
\begin{equation}
    \beta = \frac{\alpha_{\text{FC}} (F_P - 1)}{\alpha_{\text{FC}} (F_P - 1) + \dfrac{1}{\mathrm{QE}}}
    = \boxed{ \dfrac{1}{1 + \left[ \mathrm{QE}\cdot\alpha_{\text{FC}} (F_P - 1) \right]^{-1}}}.
    \label{eq:beta_SI}
\end{equation}
This expression uses the same notation as in \cite{Wang2019TurningSystem} and reduces to the familiar two-level result $\beta_F = F/(F+1)$ in the limit $\alpha_{\text{FC}} = \mathrm{QE} = 1$, upon identifying the additive enhancement factor $F = F_P - 1$ -- the convention comparison already noted for $F_P$ itself (section~\ref{subsec:Franck-Condon}). In the limit $F_P \to 1$ (no cavity), $\beta \to 0$ as expected, since no emission is added to the cavity mode; for $F_P \to \infty$, $\beta \to 1$. The overall probability that a single excitation results in a collected 00ZPL photon is given by the product $\eta_\text{ZPL} = \alpha' \cdot \beta$ (Eq.~\ref{eq:eta_zpl}).

\section{Further information on the Fabry--Perot cavity}
\label{sec:SI_FP}

\subsection{Gaussian Optics analysis for the Fabry--Perot cavity}

A Fabry--Perot microcavity is an optical resonator designed to confine and shape light with a footprint in the order of a few micrometers. It consists of a combination of reflective surfaces separated by some distance $h$~\cite{Kavokin2017Microcavities}. The supported modes are those for which the round trip $2h$ is a multiple of the wavelength $\lambda$~\cite{Saleh2019FundamentalsPhotonics}. The fundamental mode inside the microcavity can be described as a Gaussian beam that fits the mirror geometry:
\begin{equation}
I(x, y, z) = I_0 \left[\frac{w_0}{w(z)}\right]^2\, \exp \left( -2 \frac{x^2 + y^2}{w^2(z)} \right).
\end{equation}
As described in the main text, the type of resonator described here is known as a plano-concave resonator. A stable cavity consisting of a plane mirror ($R_1 = \infty$) and a concave mirror ($R_2 = R$), separated by a cavity length $h$.

The beam width $w(z)$ and the beam waist $w_0$ (at $z = z_0$) are given by~\cite{Chang2015PrinciplesAnalysis}:
\begin{equation}
w(z) = w_0 \sqrt{1 + \left( \frac{z-z_0}{z_R} \right)^2 },
\end{equation}
where the Rayleigh length is ~\cite{Chang2015PrinciplesAnalysis}:
\begin{equation}
z_R = \frac{\pi n w_0^2}{\lambda_0},
\end{equation}
and
\begin{equation}
w_0 = \left( \frac{\lambda_0}{\pi n} \right)^{1/2} \left[ h(R - h) \right]^{1/4}.
\end{equation}
Here, $z_0$ is the position of the beam waist (equal to 0 if the waist is at the plane mirror), $h$ is the cavity distance, $R$ is the concave mirror radius ($h < R$) and $\lambda_0$ is the vacuum wavelength.

The divergence angle for light coming out of the cavity is ~\cite{Chang2015PrinciplesAnalysis}:
\begin{equation}
    \theta_{\mathrm{div}} = \frac{\lambda_0}{\pi w_0}
    \qquad \text{and thus} \qquad
    \mathrm{NA} =n \sin(\theta_{\mathrm{div}}).
\end{equation}
Moreover, this Gaussian formulation allows the calculation of the effective mode volume via the effective mode length inside the resonator, quantifying the cavity effect on mode confinement~\cite{Siegman1986Lasers}:
\begin{equation}
h_{\mathrm{eff}}
=\frac{1}{2} \int_{0}^{h} \frac{I(z)}{I_0}\,dz
= \frac{1}{2} \int_{0}^{h} \frac{w_0^2}{w^2(z)}\,dz
= \frac{z_R}{2} \left[
         \arctan\left(\frac{h - z_0}{z_R}\right)
       + \arctan\left(\frac{z_0}{z_R}\right)
         \right].
\end{equation}
With this expression, it is easy to calculate the effective mode volume $V_{\mathrm{eff}}$ that is important for the calculation of the Purcell factor~\cite{Siegman1986Lasers}:
\begin{equation}
V_{\mathrm{eff}} = \frac{\pi}{2} w_0^2 h_{\mathrm{eff}}.
\end{equation}
And for completeness, the quality factor $Q$ is a dimensionless parameter that quantifies how long light remains stored in the cavity, relative to the optical period $\frac{1}{\nu_0}$~\cite{Saleh2019FundamentalsPhotonics}. For a resonator of finesse $F$ and free spectral range $\nu_\mathrm{FSR}$, it can be expressed as:
\begin{equation}
Q = \frac{\nu_0}{\nu_\mathrm{FSR}}\mathcal{F}
  = \frac{\mathrm{Re}(\omega_{\mathrm{0}})}{2\,\mathrm{Im}(\omega_{\mathrm{0}})}.
\label{eq:Qfactor}
\end{equation}

\subsection{2D Simulation and mesh definition}

FEM simulation and optimization process of the Fabry Perot cavity: The meshing and parametrization were performed in \textsc{JCMsuite}, version 6.6.5, using a parametrized layout with the \textit{multilayer} and \textit{ring} geometric operations for the bottom and top DBR, respectively. This results in a computationally manageable simulation, with each optimization step taking approximately $\SI{5}{\minute}$ using a laptop with an Intel i5-1335U microprocessor.

\begin{figure}[htbp]
    \centering
    \begin{subfigure}[b]{0.48\linewidth}
        \centering
        \includegraphics[width=\linewidth]{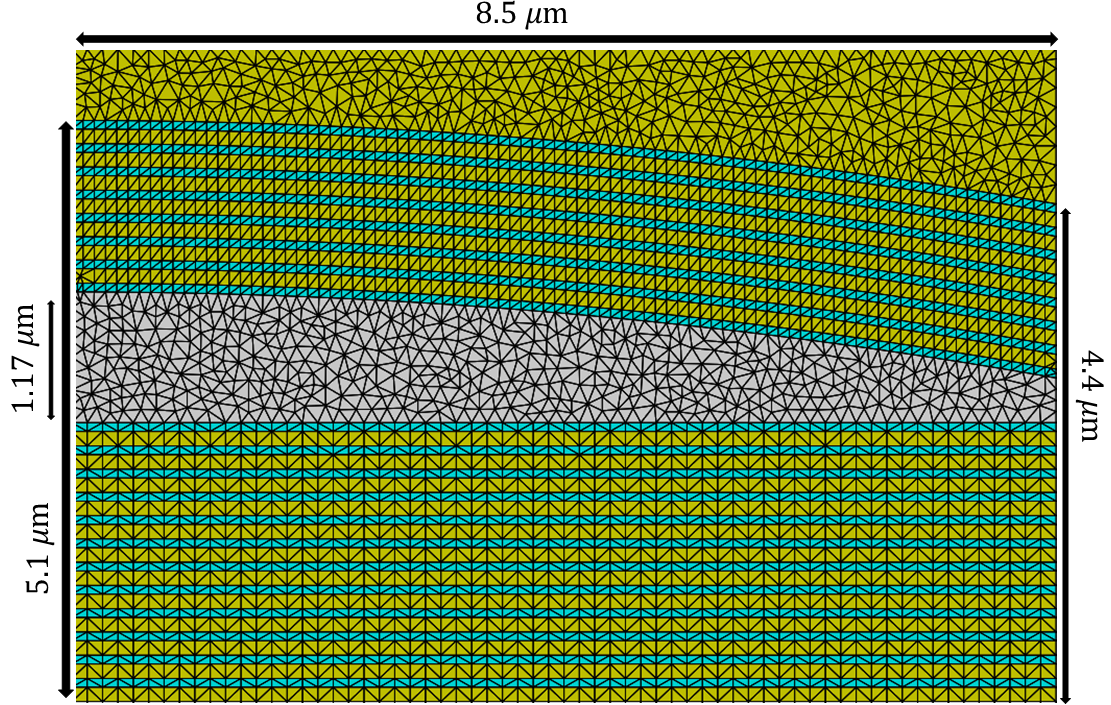}
        \caption{}
        \label{fig:mesh}
    \end{subfigure}
    \hfill
    \begin{subfigure}[b]{0.48\linewidth}
        \centering
        \includegraphics[width=\linewidth]{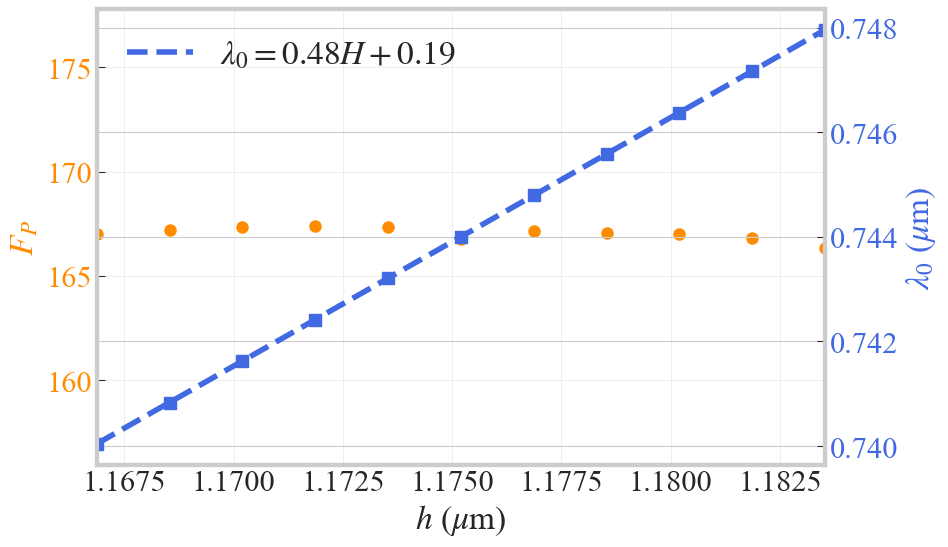}
        \caption{}
        \label{fig:fp_optimization}
    \end{subfigure}
    \caption{\textbf{FEM implementation and length sensitivity of the Fabry--Perot cavity.} \textbf{a} 2D side-view of the cavity mesh in the FEM simulator. Thinner blue layers of the DBR correspond to \ce{TiO2} while thicker green layers correspond to \ce{SiO2}. The central area in gray corresponds to \acrshort{pdcb}. The mesh elements are shown and the dimensions of the microcavity are indicated with arrows. \textbf{b} Cavity length sweep: small variations in the cavity length $h$ affect both the Purcell enhancement $F_P$ and the fundamental mode resonance wavelength $\lambda_0$, shown with a linear fit.}
    \label{fig:FP_mesh_and_sweep}
\end{figure}

As shown in Fig.~\ref{fig:mesh}, given that the spatial extent of the cavity is on the order of tens of micrometers, the meshing resolution is chosen such that there are at least three mesh elements per material wavelength. Computational efficiency could be further improved by reducing the width of the cavity in the FEM simulator to the region where the mode is confined (see Fig.~\ref{fig:Fabry-Perot}c). In this particular case, the width of the mesh becomes approximately $\SI{3}{\micro\meter}$, resulting in significantly fewer mesh cells compared to the width of $\SI{8.5}{\micro\meter}$ below.

The target of the optimization process was to maximize the free-space collection efficiency $\eta$ while pursuing a Gaussian-profiled Purcell enhancement $F_P>20$ in the emission wavelength $\lambda \approx \SI{744}{\nano\metre}$. The resonance wavelength can be tuned linearly by changing the cavity length, as shown in Fig.~\ref{fig:fp_optimization}.

\section{Further information on the micropillar cavity}
\label{sec:SI_micropillar}

\begin{figure}[htbp]
    \centering
    \includegraphics[width=\textwidth]{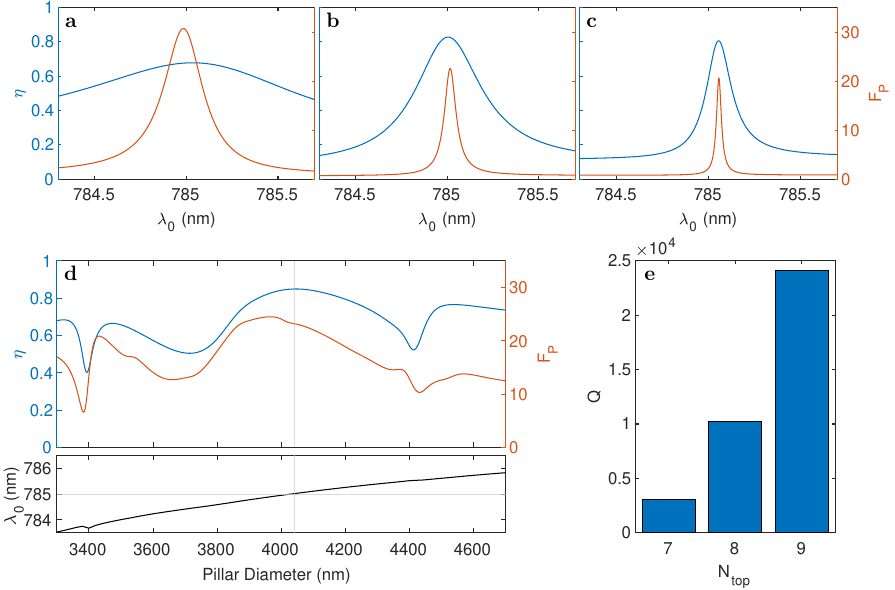}
    \caption{\textbf{Micropillar analysis:} \textbf{a-c} Collection efficiency and Purcell enhancement of structures with 7, 8 and 9 DBR pairs. The corresponding diameters are \SI{1728}{\nano\meter}, \SI{4040}{\nano\meter} and \SI{6268}{\nano\meter}, respectively. \textbf{d} The figures of merit as a function of the pillar diameter. \textbf{e} The Q-factors corresponding to the structures in \textbf{a-c}.}
    \label{fig:pillar_analysis}
\end{figure}

Similar to the Fabry--Perot cavity, the most sensitive parameter is the height of the cavity. In order to decrease the parameter space for the optimization, the number of maxima between the DBR-mirrors has been fixed, and parameter ranges for the height have been chosen accordingly. Here, we present the results for 6 maxima, including the maximum within the sublimated crystal. Figure~\ref{fig:pillar_analysis}a shows a broad enhancement, resulting from the low $Q$-factor of this configuration. In panel c additional DBR layers increase the $Q$-factor compared to panel a by almost an order of magnitude and the aspect ratio minimizes losses to the sides. The spectrum in panel b corresponds to the configuration presented in the main text.

Figure~\ref{fig:pillar_analysis}d shows the dependence of the structure on the diameter. While the center of the peak $\lambda_0$ changes slowly with the diameter, significant changes are observed in the other figures of merit that result from interferences of the fundamental Fabry--Perot mode, depicted in Fig.~\ref{fig:micropillar}c, with higher order modes. In addition to losses due to parasitic modes, the extended geometry in the bottom leads to losses to the side. Ideally, one would choose a material for the central layer of the cavity whose refractive index contrast to the sublimated crystal is as small as possible. 
\label{sec:SI_MP}

\section{Further information on the circular Bragg grating cavity}
\label{sec:SI_CBG}

\subsection{Analytical approach to the circular Bragg grating}

A \acrshort{cbg} confines an optical mode within a central disk and scatters it vertically with high efficiency~\cite{Davanco2011AEmission}. It is the cylindrical analog of a 1D grating coupler~\cite{Hardy1989AnalysisGratings, Taillaert2002AnFibers}, but designed to convert a radial in-plane mode into a mode normal to the surface, while maintaining a small mode volume for Purcell enhancement~\cite{Ates2012BrightMicrocavity}. As explained in the main text, we developed a first order analytical approach to guide our FEM simulated structures and their parameter value spaces.

\subsubsection{Phase-Matching in Cylindrical Coordinates}

For a cylindrical grating with a radial period $p$, the phase-matching condition between the radial propagation constant $\beta_r = k_0 n_{\mathrm{eff,r}}$ and a radiated wave at angle $\theta$ is
\begin{equation}
    \beta_r - k_0 \sin\theta = m \frac{2\pi}{p}, \quad \Rightarrow \quad n_{\mathrm{eff,r}} = m\frac{\lambda_0}{p} + \sin\theta.
\end{equation}
For first-order ($m=1$) vertical emission ($\theta=0$)~\cite{Taillaert2006GratingWaveguides}:
\begin{equation}
    k_0 n_{\mathrm{eff,r}} = \frac{2\pi}{p}.
    \label{eq:grating_outcoupler}
\end{equation}
This differs from the DBR quarter-wave reflectance condition $n_{\mathrm{eff}} = \lambda_0/(2p)$.

\begin{figure}[ht!]
    \centering
    \includegraphics[width=0.7\linewidth]{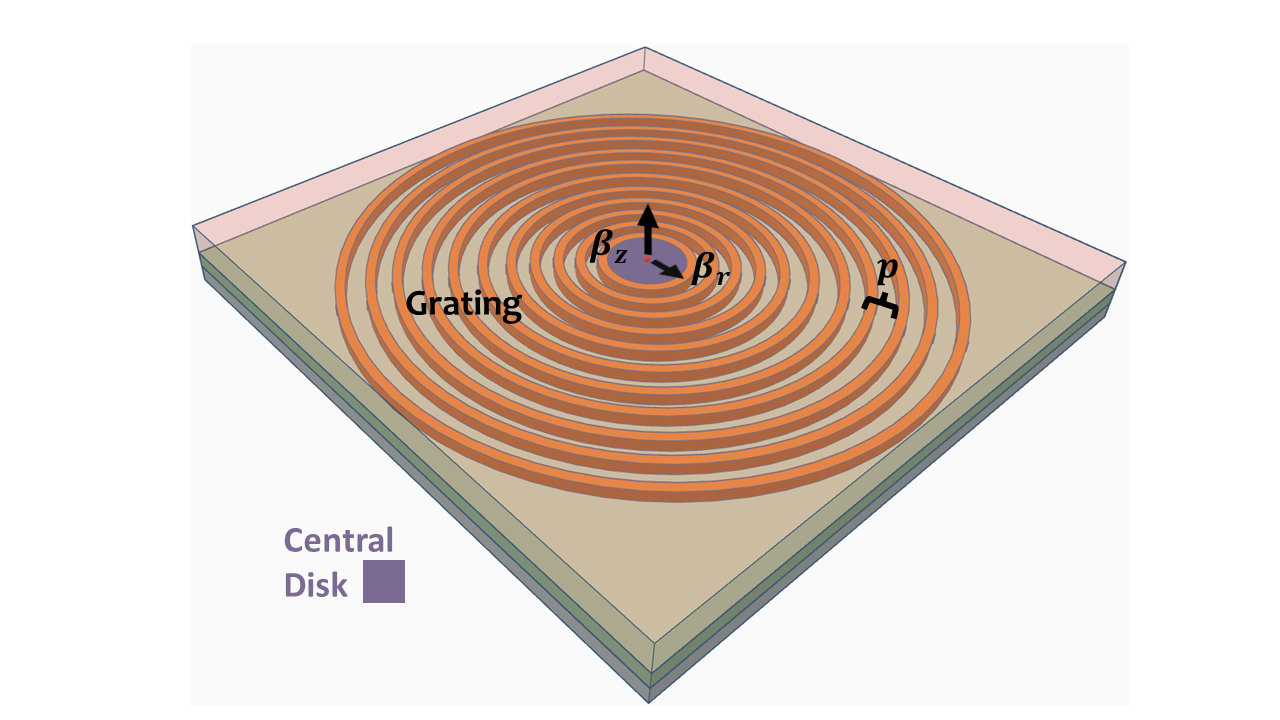}
    \caption{\textbf{Schematic of a \acrshort{cbg} as a cylindrical-grating out-coupler.} The central disk and the concentric rings form a grating. The parameters for an out-coupler are indicated: radial period $p$ and radial and vertical propagation constants $\beta_r,\beta_z$.}
    \label{fig:cbg_coupler}
\end{figure}

\subsubsection{Bloch Wave in the grating region}

A \acrshort{cbg} consists of a central defect disk, containing the emitter, surrounded by concentric Bragg rings. In the grating region, the horizontal (radial) Bloch wave leaks vertically with a constant attenuation rate for a uniform grating. This is described by the complex radial Bloch wave-vector~\cite{Joannopoulos2011PhotonicLight, Fan1998ChannelCrystals}:
\begin{equation}
    k_{r}^{\mathrm{grating}} = \frac{2\pi}{p} + i \alpha,
\end{equation}
where the real part matches the above out-coupling condition and the imaginary part $\alpha$ gives the amplitude attenuation per unit radius. In the grating region, the radial field can be represented as a real-valued attenuated Bloch wave,
\begin{equation}
    E_{\mathrm{grating}}(r)
    \propto
    e^{-\alpha (r-R_{\mathrm{disk}})}
    \cos\left[
    \frac{2\pi}{p}(r-R_{\mathrm{disk}})+\varphi
    \right],
    \qquad r \geq R_{\mathrm{disk}} .
    \label{eq:Evanescent}
\end{equation}
Here, the cosine term represents the real part of the Bloch oscillation with radial wavevector $\pi/p$, while the exponential term gives the leakage envelope. Thus,
\begin{equation}
    |E_{\mathrm{grating}}(r)|_{\mathrm{env}}
    \propto
    e^{-\alpha (r-R_{\mathrm{disk}})},
\end{equation}
so that each period reduces the field amplitude by $e^{-\alpha p}$ and the intensity by $e^{-2\alpha p}$.

Physically, satisfying the radial Bragg condition ensures that the concentric grating rings scatter the in‑plane (radial) field into the vertical ($\theta = 0$) direction. Because the period $p$ is chosen so that the scattered contributions are phase‑matched on the central axis, the upward‑scattered fields from all rings add constructively above the disk. The imaginary part $\alpha$ of the Bloch wavevector in equation~\eqref{eq:Evanescent} causes an exponential attenuation of the radial field, so most of the scattering occurs in the first rings close to $R_{\mathrm{disk}}$. As a result, when observed externally, the emission appears to originate from the center of the \acrshort{cbg}, although it is the collective interference of all grating rings that shapes the vertically directed beam. For this reason, the number of rings in the \acrshort{cbg} is an important parameter.

\subsubsection{Mode in the central disk}

Inside the uniform-index disk, the scalar Helmholtz equation in cylindrical coordinates is~\cite{Saleh2019FundamentalsPhotonics}:
\begin{equation}
    \frac{\partial^{2} E}{\partial r^{2}}
    + \frac{1}{r} \frac{\partial E}{\partial r}
    + \frac{1}{r^{2}} \frac{\partial^{2} E}{\partial \phi^{2}}
    + \frac{\partial^{2} E}{\partial z^{2}}
    + k^{2} E = 0.
\end{equation}
With $m=0$ (rotational symmetry) and the vertical propagation constant $\beta_z$, the radial equation becomes
\begin{equation}
    \frac{\mathrm{d}^{2} R}{\mathrm{d}r^{2}} 
    + \frac{1}{r} \frac{\mathrm{d}R}{\mathrm{d}r}
    + \left[ k_{r, \mathrm{disk}}^2 - \frac{m^2}{r^2} \right] R = 0,
\end{equation}
where $k_{r, \mathrm{disk}}^2 = (n k_0)^2 - \beta_z^2$ is constant in the disk. We assume pure radial in-plane propagation, therefore $\beta_z^2=0$.
For $m=0$, this is a Bessel’s equation of order zero, where the finite-at-the-origin solution is
\begin{equation}
    R(r) = A\, J_0(k_{r, \mathrm{disk}} r).
\label{eq:Bessel}
\end{equation}
$A$ is the mode amplitude and $J_0$ is the zeroth‑order Bessel function of the first kind, representing the radial oscillation of the field inside the central disk.

\subsubsection{Disk--grating matching}

At $r = R_\mathrm{disk}$, $E$ and $\partial_r E$ must be continuous:
\begin{equation}
    \frac{k_{r,\mathrm{disk}} J_1(k_{r,\mathrm{disk}} R_\mathrm{disk})}
         {J_0(k_{r,\mathrm{disk}} R_\mathrm{disk})}
    = \alpha
    \quad (\varphi=0).
    \label{eq:Match}
\end{equation}
This links the radius of the disk to the leakage rate $\alpha$. As depicted in Fig.~\ref{fig:CBG_theory}a, only certain radii yield optimal out-coupling. However, $\alpha$ cannot be obtained from this simplified model, since it depends on material properties, layer thicknesses, and other parameters of the full \acrshort{cbg} stack that are not included here; a quantitative value requires the complete FEM simulation reported in the main text. To estimate an order-of-magnitude value for illustration, we assume that the leaky mode is confined over a length comparable to the disk radius itself, i.e.\ $1/\alpha \sim R_\mathrm{disk}$. For $R_\mathrm{disk}$ of a few hundred nanometers, this gives $\alpha \approx \SI{2}{\micro\metre}^{-1}$, the value used in the plots below.

\subsubsection{Far-field pattern}

In the Fraunhofer regime, the circularly symmetric far-field is given by the zeroth-order Hankel transform of the aperture envelope~\cite{Goodman2005IntroductionOptics}:
\begin{equation}
    E_{\mathrm{far}}(k_\perp) \propto 
    \int_{0}^{\infty} e^{-\alpha r} J_{0}(k_\perp r)\, r\, \mathrm{d}r
    = \frac{\alpha}{\left(\alpha^{2} + k_\perp^{2}\right)^{3/2}},
\end{equation}
with intensity
\begin{equation}
    I_{\mathrm{far}}(\theta) \propto
    \frac{\alpha^2}{\left[\alpha^2 + k_0^2\sin^2\theta\right]^3}.
    \label{eq:Far_field}
\end{equation}
From this we can see in Fig.~\ref{fig:CBG_theory}c an exponential envelope that yields a Gaussian-like far-field pattern. For future work, apodization could aim for a Gaussian far-field for better fiber coupling ~\cite{Taillaert2006GratingWaveguides, Hekmati2023BullseyeQuantum-light-emitter}.

\subsubsection{Mode volume}

\begin{figure}[ht!]
    \centering
    \includegraphics[width=\linewidth]{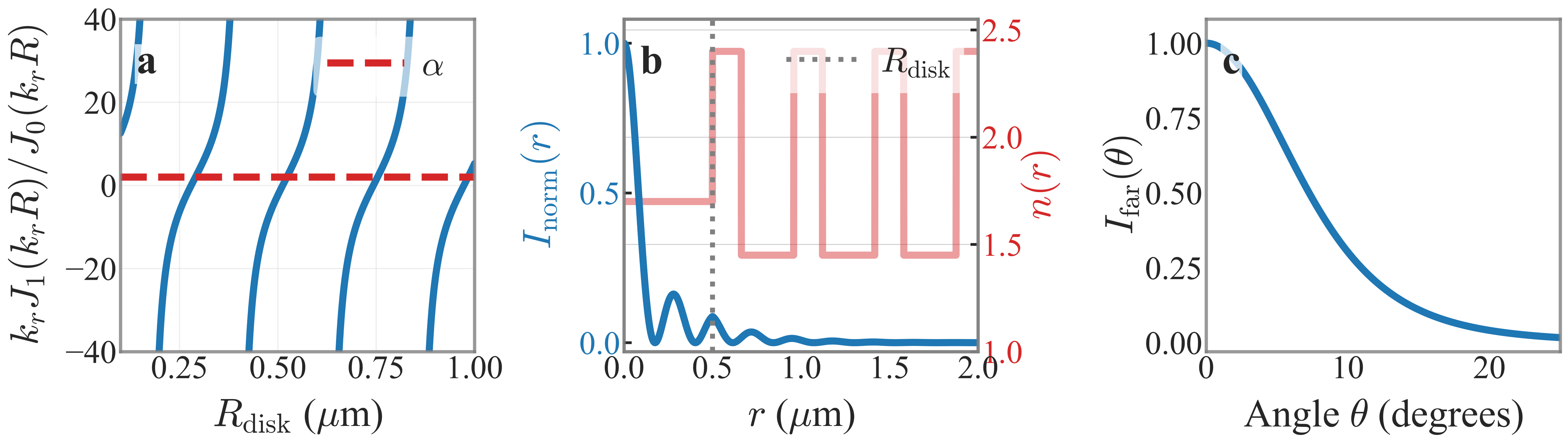}
    \caption{
    \textbf{Expected analytical behavior of an optimized circular Bragg grating.}
    \textbf{a} Theoretical \acrshort{cbg} matching condition (Eq.~\ref{eq:Match}) for $\alpha= \SI{2}{\micro\meter}^{-1}$, $\lambda=\SI{0.78}{\micro\meter}$ and $n=1.7$. Only certain values for the central disk radius will result in a resonant microcavity. 
    \textbf{b} Normalized radial field intensity profile, combining the disk mode (Eq.~\ref{eq:Bessel}, $J_0$ oscillations for $r<R_\mathrm{disk}$) with the attenuated Bloch-wave in the grating region (Eq.~\ref{eq:Evanescent}), for $R_\mathrm{disk} = \SI{0.5}{\micro\meter}$ and $\alpha = \SI{2}{\micro\meter}^{-1}$. A refractive index profile of alternating \ce{TiO2}/\ce{SiO2} rings is overlaid for reference.
    \textbf{c} Plot of equation~\ref{eq:Far_field}. Far‑field intensity vs. collection angle, the shape is close to a Gaussian.}
    
    \label{fig:CBG_theory}
\end{figure}

The effective mode volume is given by:
\begin{equation}
    V_{\mathrm{eff}} =
    \frac{\int \varepsilon |E|^2\,dV}{\varepsilon_\mathrm{max} |E_\mathrm{max}|^2}.
\end{equation}
For a uniform \acrshort{cbg}, $|E(r)|^2$ decays as $e^{-2\alpha (r - R_\mathrm{disk})}$ in the grating, so most of the energy is confined within $\approx 1/\alpha$ (see Fig.~\ref{fig:CBG_theory}b). Considering $\alpha \approx \SI{2}{\micro\metre}^{-1}$, $V_{\mathrm{eff}}$ is on the order of $(\lambda/n)^3$. Although the leakage rate $\alpha$ is not directly experimentally accessible for a \acrshort{cbg} cavity, the analytical treatment above -- the Bessel-like disk mode, the exponential leakage into the grating, and the resulting scaling of the mode volume -- reproduces the qualitative behavior expected of a \acrshort{cbg} cavity and agrees with the design principles established for quantum-dot \acrshort{cbg} cavities~\cite{Davanco2011AEmission, Ates2012BrightMicrocavity, Rickert2023High-performanceGratings, Rickert2025AClock-rates}. This analytical picture therefore provides physical intuition that complements the quantitative FEM simulations reported in the main text.

\subsection{CBG Optimization Process}

In order to keep the parameter space for optimization small, only three parameters are free. The disk radius $R_{\mathrm{disk}} \in [100, 600]\,\si{\nano\metre}$, the period width $p\in [375, 575]\,\si{\nano\metre}$ and the ridge width $w\in [50, 300]\,\si{\nano\metre}$. The values are chosen following theoretical intuition (Eq.~\ref{eq:grating_outcoupler}) and reference \cite{Rickert2023High-performanceGratings}, where the low computational cost of the algorithm was essential to explore more extreme parameter values than those near the out-coupler condition, initially studied in reference \cite{Davanco2011AEmission}. The target of the optimization process was to maximize the efficiency of free space collection $\eta$ while pursuing a Purcell enhancement above $20$ in the emission wavelength $\lambda\approx\SI{785}{\nano\metre}$. As in previous work, simultaneously achieving high Purcell enhancement and collection efficiency proved challenging within the available parameter space~\cite{Ates2012BrightMicrocavity}.

\subsection{CBG Parameter Dependencies}

\begin{figure}[ht!]
    \centering
    \includegraphics[width=\textwidth]{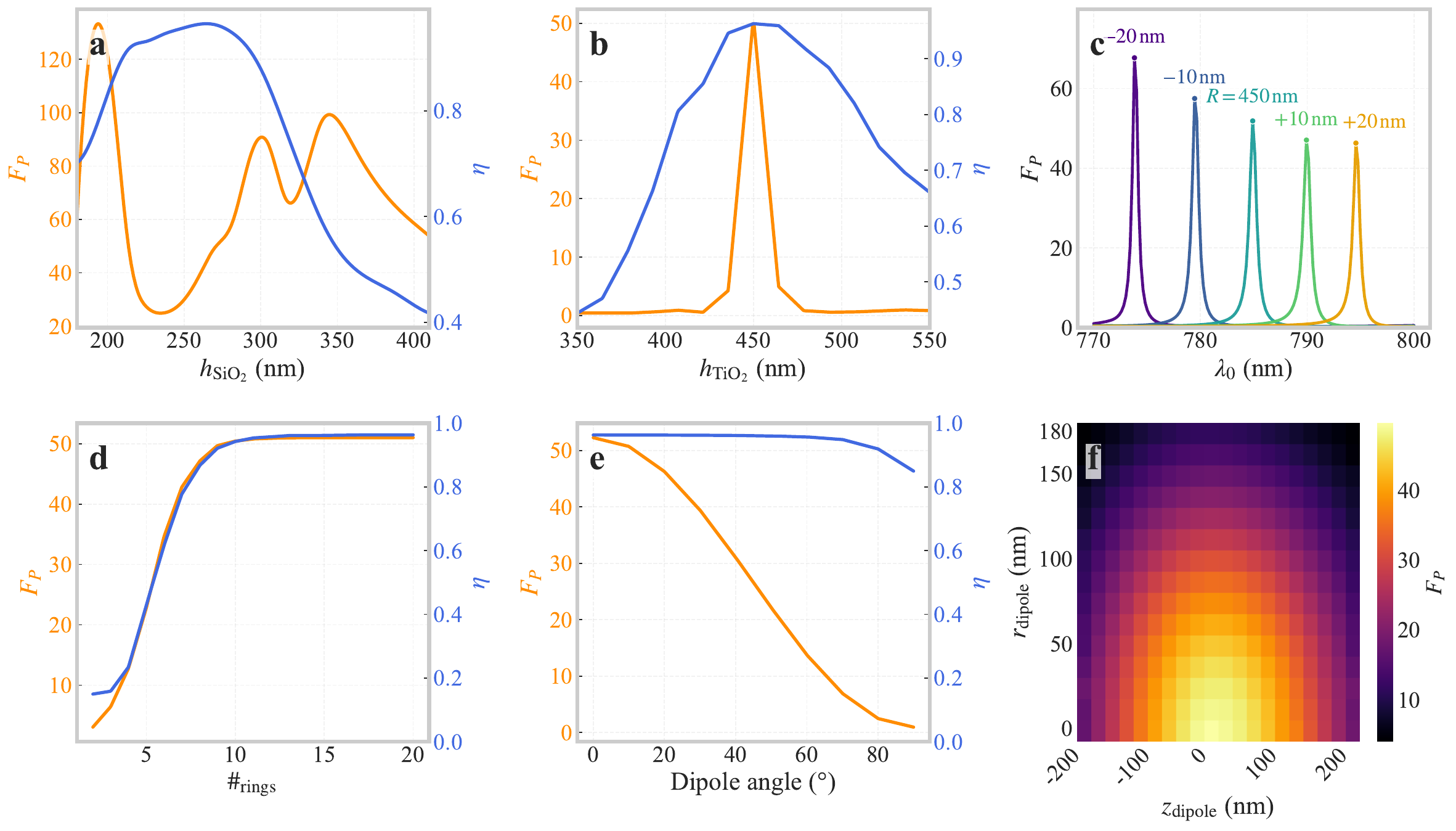}
    \caption{\textbf{Parameter dependencies for the circular Bragg grating cavity.} 
             \textbf{a} Sweep of the \ce{SiO2} spacer thickness.
             \textbf{b} Sweep of the disk height.
             \textbf{c} Sensitivity of the resonance wavelength to the central disk radius $R_\text{disk}$.
             \textbf{d} Saturation of performance with the number of rings.
             \textbf{e} Dependence on dipole orientation.
             \textbf{f} $F_P$ dependence of variations in the dipole position.}
    \label{fig:CBG_parameter_sweeps}
\end{figure}

Some other design dependencies of the \acrshort{cbg} are of interest for future fabrication and integration with molecules. In the following sweeps, values of $F_P$ and $\eta$ are taken as their maximum values within the spectral range of $\lambda_0 \pm \SI{1}{\nano\metre}$. We first investigated whether the initial, somewhat arbitrary choice of $\lambda/2$ for the \ce{SiO2} spacer thickness was optimal. As shown in the sweep in Fig.~\ref{fig:CBG_parameter_sweeps}a, a thickness of $h_{\mathrm{SiO_2}} = \SI{270}{\nano\meter}$ combines a high Purcell enhancement $F_P>50$ with the maximum of collection efficiency at the desired wavelength. This results in a well-designed cavity with a reduced parameter space, which facilitates efficient optimization. However, if one instead seeks the global maximum of $F_P$ or $\eta$ individually, treating the \ce{SiO2} height as a variable parameter (at the cost of an enlarged parameter space) may yield a different, and potentially higher, value than the balanced choice above.

We then performed a parametric sweep to determine the optimal disk thickness. As shown in Fig.~\ref{fig:CBG_parameter_sweeps}b, the Purcell factor saturates at a height of \SI{450}{\nano\metre}, identifying this as the optimal trade-off between optical performance without enlarging the computational space. This value aligns with the devices currently under experimental characterization, accounting for experimental relevance. 

As indicated in the main text, small deviations in the disk radius affect both the Purcell enhancement and the resonance wavelength. In this case, fabrication defects of a few nanometers will shift the resonance wavelength (see Fig.~\ref{fig:CBG_parameter_sweeps}c).

Moreover, two parameters that were set without discussion are investigated: the number of rings and the dipole orientation in the $xy$ plane. The performance associated with the number of Bragg rings saturates at approximately 10 rings (see Fig.~\ref{fig:CBG_parameter_sweeps}d). In Fig.~\ref{fig:CBG_parameter_sweeps}e, the Purcell factor $F_P$ shows a cosine dependence on the dipole orientation angle relative to the cavity axis, while the collection efficiency remains stable up to $\approx70^\circ$.

Finally, the dependence on emitter position was investigated, since perfect placement of the emitter in the center of the \acrshort{cbg} is in general not feasible. From Fig.~\ref{fig:CBG_parameter_sweeps}f and in agreement to other works \cite{Rickert2019OptimizedGratings}, the tolerance for this parameter seems to be on the order of tens of nanometers in both the radial and vertical directions.

\newpage
\printbibliography[title=Supporting Information References]

\end{refsection}

\end{document}